\documentclass[preprintnumbers]{revtex4}
\usepackage{amssymb}
\usepackage{makeidx}
\usepackage{amsmath}
\usepackage{graphicx}
\usepackage{dcolumn}
\usepackage{bm}

\begin{document}

\preprint{}
\title{Phase engineering of chirped rogue waves in Bose--Einstein
condensates with a variable scattering length in an expulsive potential}
\author{Emmanuel Kengne$^{1}$ }
\thanks{Corresponding author: ekengne6@zjnu.edu.cn}
\author{Boris A. Malomed$^{2,3}$}
\author{ WuMing Liu$^{4}$}
\affiliation{$^{1}$ School of Physics and Electronic Information Engineering, Zhejiang
Normal University, Jinhua 321004, China \\
$^{2}$ Department of Physical Electronics, School of Electrical Engineering,
Faculty of Engineering, and Center for Light-Matter Interaction, Tel Aviv
University, P.O.B. 39040, Ramat Aviv, Tel Aviv, Israel\ \\
$^{3}$Instituto de Alta Investigaci\'{o}n, Universidad de Tarapac\'{a},
Casilla 7D, Arica, Chile \\
$^{4}$ National Laboratory for Condensed Matter Physics, Institute of
Physics, Chinese Academy of Sciences, Beijing 100190, P.R. China\ }
\date{\today }

\begin{abstract}
We consider a cubic Gross-Pitaevskii (GP) equation governing the dynamics of
Bose-Einstein condensates (BECs) with time-dependent coefficients in front
of the cubic term and inverted parabolic potential. Under a special
condition imposed on the coefficients, a combination of phase-imprint and
modified lens-type transformations converts the GP equation into the
integrable Kundu-Eckhaus (KE) one with constant coefficients, which contains
the quintic nonlinearity and the Raman-like term producing the
self-frequency shift. The condition for the baseband modulational
instability of CW states is derived, providing the possibility of generation
of chirped rogue waves (RWs) in the underlying matter-wave (BEC) model.
Using known RW solutions of the KE equation, we present explicit first- and
second-order chirped RW states. The chirp of the first- and second-order RWs
is independent of the phase imprint. Detailed solutions are presented for
the following configurations: (i) the nonlinearity exponentially varying in
time; (ii) time-periodic modulation of the nonlinearity; (iii) a stepwise
time modulation of the strength of the expulsive potential. Singularities of
the local chirp coincide with valleys of the corresponding RWs. The results
demonstrate that the temporal modulation of the \textit{s}-wave scattering
length and strength of the inverted parabolic potential can be used to
manipulate the evolution of rogue matter waves in BEC.\bigskip \newline
\bigskip

$\mathcal{DEDICATION}$\textbf{:} The first author, \textit{E. Kengne},
dedicates this work to his brother, \textbf{Sir Philippe Wambo}
\end{abstract}

\maketitle

\section{Introduction}

Extensive research work on nonlinear dynamics of atomic matter waves in
Bose-Einstein condensates (BECs), such as dark \cite{1,2,6} bright \cite%
{6,7,8}, and breather \cite{breather1,breather2} solitons, rogue waves (RWs)
\cite{9,10}, gap solitons and other coherent modes in optical lattices \cite%
{11,Morsch}, four-wave mixing \cite{12}, and, most recently, quantum
droplets in binary \cite{Petrov,binary} and dipolar \cite{dipolar} BEC set
these topics at the forefront of current experimental and theoretical
studies in soft-matter physics and nonlinear science. A commonly adopted
dynamical model of BEC is the Gross-Pitaevskii (GP)\ equation (including the
Lee-Huang-Yang corrections that account for the beyond-mean-field effects
helping to stabilize quantum droplets \cite{Petrov}). In the ideal
one-dimensional (1D) form, which does not include losses, an external
potential, and variable coefficients, the GP equation is tantamount to the
classical integrable nonlinear Schr\"{o}dinger (NLS) equation, which
produces commonly known exact solutions for solitons and elastic
interactions between them \cite{Zakharov,21}. 
Nevertheless, some specially designed integrable models admit exact
solutions for fusion and fission of solitons \cite{15,16,we}.

Similar to other nonlinear media, BECs may support the creation of rogue
waves (RWs) \cite{Konotop}, i.e., peaks spontaneously emerging on top of an
unstable CW (continuous-wave, i.e., constant-amplitude) modulationally
unstable background, and disappearing afterwards \cite{mi1,mi2,mi3}.
Generally, rogue waves are known as waves which suddently appear in the
oceans that can reach amplitudes more than twice the value of ignificant
wave height \cite{jul1}. As far as Bose-Einstein condensates are concerned a
sudden increase of peaks in the condensate clouds is very similar to the
nature of appearance of high peaks in the open ocean \cite{jul1a}.
Mathematically, rational solutions of some nonlinear partial differential
equations such as the NLS equation play a major role in the theory of rogue
waves \cite{jul2}. Motivated by the above observation, we aim to address in
this work, the dynamics of chirped rogue matter waves in the framework of
the GP equation with a time-varying atomic scattering length, which
determines the coefficient in front of the cubic term, and a time-dependent
strength of the parabolic potential. The analysis is performed with the help
of the \textit{phase-engineering technique} \cite{17,we}, which imposes a
phase imprint onto the BEC's mean-field wave function. By means of this
technique, we engineer the imprinted phase which makes it possible to
transform the cubic GP equation into one including the cubic-quintic
nonlinearity and a self-frequency-shift term. In particular, the quintic
term may account for effects of three-body collisions in BEC \cite%
{20,quintic,17}, as well as higher-order nonlinearity in optics \cite%
{29,Reyna}.

In the physically important case of the cigar-shaped BECs, the GP equation
reduces to the 1D NLS equation with the external potential \cite{6,18}:

\begin{equation}
i\frac{\partial \psi (x,t)}{\partial t}+\frac{\partial ^{2}\psi (x,t)}{%
\partial x^{2}}+2a(t)\left\vert \psi (x,t)\right\vert ^{2}\psi
(x,t)-k(t)x^{2}\psi (x,t)=0.  \label{1}
\end{equation}%
Here, $\psi (x,t)$ is the normalized mean-field wave function, time $t$ and
coordinate $x$ are measured in units $2/\omega _{\bot }$ and $a_{\bot }$,
where $a_{\bot }=\sqrt{\hslash /(m\omega _{\bot })}$ is the
harmonic-oscillator lengths of the transverse confining potential with
frequency $\omega _{\perp }$, $m$ is the atomic mass, and $\omega _{0}$ is
the frequency of the longitudinal trapping potential, $k=2\omega
_{0}^{2}/\omega _{\bot }^{2}\ll 1$ being its relative strength, which may be
a time-dependent parameter \cite{we}. The nonlinearity coefficient may also
be made a function of time by means of the Feshbach-resonance (FR) \textit{%
management}, if FR is imposed by a variable magnetic filed \cite{book}. In
this case, the nonlinearity coefficient in Eq. (\ref{1}) is $a(t)=\left\vert
a_{s}(t)\right\vert /a_{B}$, where $a_{B}$ is the Bohr radius, and $%
a_{s}(t)<0$ is the \textit{s}-wave scattering length of collisions between
attractively interacting atoms. In the experiment, bright solitons were
created by utilizing FR to switch the sign of the \textit{s}-wave scattering
length from positive to negative \cite{8} values. In the GP equation (\ref{1}%
), the 1D wave function $\psi (x,t)$ is related to the original 3D one, $%
\Psi (\mathbf{r},t)$, by expression
\begin{equation}
\Psi (\mathbf{r},t)=\frac{1}{\sqrt{2\pi a_{B}a_{\bot }}}\psi \left( \frac{x}{%
a_{\bot }},\frac{\omega _{\bot }t}{2}\right) \exp \left( -i\omega _{\bot }t-%
\frac{x^{2}+y^{2}}{2a_{\bot }}\right) .
\end{equation}

The 3D and 1D equations are very different in terms of their stability. In
the true 1D system, the collapse does not occur with the increase of the
number of atoms. However, the realistic 1D limit is not tantamount to the
ideal NLS equation, the deviation from which makes the collapse possible
\cite{NJP,breather2}. Nevertheless, in Ref. \cite{7} it was demonstrated
that, in a safe range of parameters, one can avoid the collapse of the
condensates, while exponentially increasing $a(t)$ by means of the FR
management. The repulsive three-body interatomic interactions, which are
represented by the above-mentioned quintic term added to the GP equation,
can also enhance the stability of BEC \cite{20,quintic,17}. The inclusion of
the latter term allows one to generate high-density BEC, while retaining the
one-dimensionality of the system without severely restricting to the
parametric domain.

Parameter $a(t)$ of the two-body interatomic interaction being
time-dependent, Eq. (\ref{1}) can be used to describe the control and
management of BEC \cite{book,we}. As concerns a possibility to design an
integrable version of Eq. (\ref{1}), Kumar \textit{et al}. \cite{17} had
produced a Lax pair associated with a specific form of the cubic-quintic GP
equation, and had thus constructed bright solitons employing a
gauge-transformation method. It was thus demonstrated that, for attractive
three-body interactions, solitons of the cubic-quintic GP equation with an
exponentially increasing scattering length and constant potential strength
differ from those of the cubic GP equation by an additional phase, while the
density of the condensates remains the same.

Proceeding to chirped solitons, it is relevant to mention that, because of
their ability to produce very narrow outputs, chirped pulses are
particularly useful in photonics, for the design of fiber-optic amplifiers,
optical pulse compressors, and soliton-based communications links \cite{22}.
Pulses with the linear chirp and a hyperbolic-secant-amplitude profile were
investigated numerically and analytically \cite{23}. The existence of
chirped soliton-like solutions of the cubic-quintic NLS equation without an
external potential was reported too \cite{24}. Within the framework of a
generalized NLS equation containing the group-velocity dispersion, Kerr and
quintic nonlinearity, and self-steepening effect, Chen \textit{et al}. \cite%
{25} had investigated the super-chirped RW dynamics in optical fibers, using
the nonrecursive Darboux-transformation technique. Most recently, Mouassom
\textit{et al}. \cite{26} have combined the similarity transformation and
the use of a direct ansatz to solve analytically an inhomogeneous chiral NLS
equation with modulated coefficients. They have also investigated the RW
propagation with a chirped structure. The existence of chirped solitons in
BEC models with in external potentials was reported too \cite{27}.

The main purpose of this work is to map the cubic GP Eq. (\ref{1}) into an
integrable cubic-quintic NLS equation including a self-frequency shift term,
which is known as the Kundu-Eckhaus equation \cite{300}. Using solutions of
the KE equation makes it possible to generate chirped RWs in BEC with both
two- and three-body interatomic interactions in the external
harmonic-oscillator trap by suitably engineering the phase imprint imposed
on the wave function governed by Eq. (\ref{1}). The rest of the work is
organized as follows. In Section II, we combine the phase engineering
technique with a modified lens-type transformation to map the cubic GP Eq. (%
\ref{1}) into an integrable cubic-quintic NLS equation with a self-frequency
shift term, which is then used to produce analytical chirped RW solutions
for BEC with the time-varying atomic scattering length in an external
parabolic potential. In Section III, the obtained exact solutions are used
to investigate the effects of the three-body interatomic interactions on the
chirped RWs. Main results of the work are summarized in Section IV.

\section{Phase engineering and chirped solutions}

\subsection{The transformations}

To derive analytical chirped-wave solutions of Eq. (\ref{1}), we combine the
phase-engineering technique \cite{17} and a modified lens-type
transformation, by means of the following ansatz:
\begin{equation}
\psi (x,t)=\frac{1}{\sqrt{\ell }}\Phi (X,T)\exp \left[ i\left\{ \alpha
(t)x^{2}-2\theta (x,t)\right\} \right] ,  \label{2}
\end{equation}%
where $\Phi (X,T)$ is a new complex wave function, $T=T(t)$, $\ell =\ell (t)$
and $\alpha (t)$ are real functions of time $t$, $X\equiv x/\ell (t)$, and $%
\theta (x,t)$ is the phase imprint, related to $\Phi (X,T)$ as follows:
\begin{subequations}
\begin{eqnarray}
\frac{\partial \theta }{\partial x} &=&-\frac{a_{3}}{4\ell }\left\vert \Phi
\right\vert ^{2},  \label{3a} \\
\frac{\partial \theta }{\partial t} &=&i\frac{a_{3}}{4}\left[ \frac{1}{\ell
^{2}}\left( \Phi \frac{\partial \Phi ^{\ast }}{\partial X}-\Phi ^{\ast }%
\frac{\partial \Phi }{\partial X}\right) -4ix\alpha (t)\frac{1}{\ell }%
\left\vert \Phi \right\vert ^{2}\right] +\frac{\allowbreak 4a_{1}+a_{3}^{2}}{%
8\ell ^{2}}\left\vert \Phi \right\vert ^{4}.  \label{3b}
\end{eqnarray}%
Here, $a_{1}$ and $a_{3}$ are two real parameters of the phase engineering
technique. In the special case when $\alpha (t)=0$ and $\ell (t)=1$, the
substitution of ansatz (\ref{2}) with the phase imprint defined by Eqs. (\ref%
{3a}) and (\ref{3b}) transforms the cubic GP Eq. (\ref{1}) into a
cubic-quintic GP equation with a cubic derivative term \cite{17}.

In this work, we focus on the general situation with $\alpha (t)\neq 0$. We
then demand that
\end{subequations}
\begin{subequations}
\begin{eqnarray}
\frac{dT}{dt} &=&\frac{1}{\ell ^{2}},  \label{4b} \\
\frac{1}{\ell }\frac{d\ell }{dt}-4\alpha &=&0,  \label{4c} \\
\frac{d\alpha }{dt}+4\alpha ^{2}+k(t) &=&0.  \label{4d}
\end{eqnarray}%
The choice of Eq. (\ref{4b}) is made to preserve the scaling. Then, Eq. (\ref%
{1}) in terms of rescaled variables $X$ and $T$ is converted into the
following cubic-quintic NLS equation with an additional cubic derivative
term:
\end{subequations}
\begin{equation}
i\frac{\partial \Phi }{\partial T}+\frac{\partial ^{2}\Phi }{\partial X^{2}}%
+2\ell a(t)\left\vert \Phi \right\vert ^{2}\Phi +\allowbreak a_{1}\left\vert
\Phi \right\vert ^{4}\Phi +ia_{3}\left( \frac{\partial }{\partial X}%
\left\vert \Phi \right\vert ^{2}\right) \Phi =0,  \label{5}
\end{equation}%
where $a_{1}$ and $2\ell a(t)>0$ represent the quintic and cubic
nonlinearities, respectively, and $a_{3}$ is the self-frequency shift
coefficient. In the context of fiber optics, Eq. (\ref{5}) may be used to
model the propagation of ultrashort pulses in a single-mode optical fiber
\cite{28}. In that context, $T$ and $X$ are the propagation distance and
retarded time, respectively, $2\ell a(t)$ is the coefficient of the cubic
nonlinearity [in the optics counterpart of Eq. (5), this coefficient may
then be rescaled into one multiplying the group-velocity dispersion (GVD)
term, which is termed anomalous and normal GVD for $a(t)>0$ and $a(t)<0$,
respectively] \cite{29,25}, and real $a_{3}$ is related to the fiber's
nonlinearity dispersion. In the context of BECs, Eq. (\ref{5}) may be used
as the GP equation including both two- and three-body interatomic
interactions, with $2\ell a(t)$ and $a_{1}$ representing the strengths of
these interactions, respectively. As mentioned by Kumar \textit{et al}. in
Ref. \cite{17}, this way of engineering the phase imprint to generate a new
integrable model is reminiscent of generating dark solitons in BEC\ by dint
of phase imprinting \cite{29a}.

To provide integrability of Eq. (\ref{5}) and exploit some known results of
the integrable KE equation \cite{300}, we adopt the linkage of strength $%
a_{1}$ of the three-body interactions to the self-frequency-shift parameter $%
a_{3}$, \textit{viz}.,
\begin{equation}
a_{1}=a_{3}^{2}/4,  \label{1/4}
\end{equation}%
and impose condition
\begin{subequations}
\begin{equation}
\ell (t)=\frac{a_{2}}{2a(t)},\text{ \ }  \label{4a}
\end{equation}%
where $a_{2}$ is an arbitrary positive real parameter. We then obtain from
Eqs. (\ref{4b})--(\ref{4d}) that the potential's and nonlinearity strengths,
$k(t)$ and $a(t)$, must satisfy a constraint,
\begin{equation}
\frac{d}{dt}\left( \frac{1}{a}\frac{da}{dt}\right) -\left( \frac{1}{a}\frac{%
da}{dt}\right) ^{2}-4k(t)=0,  \label{4e}
\end{equation}%
while $T(t)$ and $\alpha (t)$ must be defined as
\begin{equation}
T(t)=\frac{4}{a_{2}^{2}}\int_{0}^{t}a^{2}(z)dz,\text{ }\alpha (t)=-\frac{1}{%
4a}\frac{da}{dt}.  \label{4f}
\end{equation}%
Note that Eq. (\ref{4e}) is a Riccati type equation for function $%
a^{-1}da/dt $. Regardless of what $a(t)$ is, as long as condition (\ref{4e})
holds the underlying cubic GP equation (\ref{1}) is integrable. Henceforth,
we call equation (\ref{4e}) the integrability condition\ for the cubic
equation (\ref{1}) (condition (\ref{1/4}) is not by itself necessary for
selecting the integrable version of Eq. (\ref{1})). It is also important to
mention that the Painlev\'{e} singularity-structure analysis performed on
Eq. (\ref{1}) confirms that the same condition (\ref{4e}) is a necessary
integrability conditions. Finally, we note that the solution based on ansatz
(\ref{2}) provides flexibility to generate new structures related to chirped
RWs, which may be relevant for experiments with BEC.

By choosing the self-frequency shift coefficient as $a_{3}=-2\sqrt{2a_{2}}%
\beta $ and using transformation $\tau =\left( a_{2}/2\right) T,$ $\zeta =%
\sqrt{a_{2}/2}X$, Eq. (\ref{5}), subject to condition (\ref{4a}), reduces to
the integrable Kundu-Eckhaus equation \cite{300},
\end{subequations}
\begin{equation}
i\frac{\partial \Phi }{\partial \tau }+\frac{\partial ^{2}\Phi }{\partial
\zeta ^{2}}+2\left\vert \Phi \right\vert ^{2}\Phi +4\beta ^{2}\left\vert
\Phi \right\vert ^{4}\Phi -4i\beta \left( \frac{\partial }{\partial \zeta }%
\left\vert \Phi \right\vert ^{2}\right) \Phi =0,  \label{6}
\end{equation}%
which finds applications to nonlinear optics, quantum field theory, and
matter waves \cite{300,30,30a}; here, $\beta $ is a real parameter of
arbitrary sign. Most recently, Kengne and Liu in Ref. \cite{17} have derived
the cubic-quintic NLS equation (\ref{5}) to engineer RWs in a modified
Nogochi nonlinear electric transmission network \cite{31}.

As mentioned by Kumar \textit{et al}. \cite{17}, the last term in Eq. (\ref%
{6}) offsets the modulation instability driven the three-body interactions.
As pointed out in Ref. \cite{20}, strength $4\beta ^{2}$ of the three-body
interaction in Eq. (\ref{6}) is usually small in comparison to the strength
of the two-body interaction, which is represented by cubic term in Eq. (\ref%
{6}). Accordingly, we here set $2\beta ^{2}\equiv \chi _{0},$ with $0\leq
\chi _{0}<1$, hence $\beta $ takes values $\left\vert \beta \right\vert <1/%
\sqrt{2}\approx 0.70711$. Because the strength of the three-body interaction
is related to that of the binary interaction, $\beta $ may be controlled by
the tuning the \textit{s}-wave scattering length, with the help of FR.

Lastly, we note that relation (\ref{2}) implies that solutions of the
cubic-quintic NLS equation (\ref{5}) has the same density structure as
solutions of the cubic GP equation (\ref{1}), differing from them by the
phase pattern.

\subsection{Baseband modulational instability analysis}

It is known that the modulational instability (MI) is the basic mechanism
that may lead to\ the excitation of RW in nonlinear media \cite{mi1,mi2,mi3}%
. Appearing in many nonlinear dispersive systems, MI is associated with the
growth of spatially periodic perturbations added to an unstable CW\
background, and indicates that, due to the interplay between the
nonlinearity and dispersive effects, a small perturbation may lead to
breakup of the CW into a train of localized waves \cite{mi4}. Most recently,
it has been revealed that not every kind of MI necessarily leads to the
generation of RWs, and, generally, it is the baseband MI that plays such a
pivotal role in RW generation \cite{mi1,mi5}. Baseband MI implies that the
CW background may be unstable against perturbations with infinitesimally
small wavenumbers \cite{mi1}.

We begin with the CW solution of Eq. (\ref{5}) that, under condition (\ref%
{4a}), amounts to%
\begin{equation}
\Phi _{0}(X,T)=\rho _{0}\exp \left[ i\left\{ \Omega _{0}X+\left( a_{2}\rho
_{0}^{2}+\allowbreak a_{1}\rho _{0}^{4}-\Omega _{0}^{2}\right) T\right\} %
\right] ,  \label{m1}
\end{equation}%
where $\rho _{0}$ and $\Omega _{0}$ are, respectively, the amplitude and
wavenumber, while real $\widetilde{\omega }=$ $a_{2}\rho
_{0}^{2}+\allowbreak a_{1}\rho _{0}^{4}-\Omega _{0}^{2}$ is the frequency.
To address the baseband MI, we add small modulational perturbations to the
CW solution (\ref{m1}):%
\begin{equation}
\Phi (X,T)=\Phi _{0}(X,T)\left\{ 1+b_{1}\exp \left[ i\Omega \left(
X-KT\right) \right] +b_{2}^{\ast }\exp \left[ -i\Omega \left( X-K^{\ast
}T\right) \right] \right\} ,  \label{m2}
\end{equation}%
where $b_{1}$ and $b_{2}$ are two small complex amplitudes, $\Omega $ and $%
\Omega K$ are, respectively, the wavenumber and frequency of the modulation,
and $\ast $ stands for the complex conjugate. The MI emerges when $K$
becomes complex. Substituting Eq. (\ref{m2}) into Eq. (\ref{5}) under
condition (\ref{4a}) and linearizing the resulting equation with respect to $%
b_{1}$ and $b_{2}$, we obtain a linear homogeneous algebraic system for
amplitudes $b_{1}$ and $b_{2}$:%
\begin{equation}
\left\{
\begin{array}{c}
\left( K\Omega -2\Omega \Omega _{0}-\Omega ^{2}+a_{2}\rho
_{0}^{2}+2a_{1}\rho _{0}^{4}-a_{3}\rho _{0}^{2}\Omega \right) b_{1}+\left(
a_{2}\rho _{0}^{2}+2a_{1}\rho _{0}^{4}-a_{3}\rho _{0}^{2}\Omega \allowbreak
\right) b_{2}=0, \\
\left( a_{2}\rho _{0}^{2}+2a_{1}\rho _{0}^{4}+a_{3}\rho _{0}^{2}\Omega
\right) b_{1}+\left( -K\Omega +2\Omega \Omega _{0}-\Omega ^{2}+a_{2}\rho
_{0}^{2}+2a_{1}\rho _{0}^{4}+a_{3}\rho _{0}^{2}\Omega \right) b_{2}=0.%
\end{array}%
\right.
\end{equation}%
This system has a nontrivial solution if $\Omega $ and $K$ satisfy the
linear dispersion relation,%
\begin{equation}
\Omega ^{2}\left[ \allowbreak \left( \allowbreak K-2\Omega _{0}-a_{3}\rho
_{0}^{2}\right) ^{2}-\Omega ^{2}+\rho _{0}^{2}\left( \allowbreak
2a_{2}+\left( 4a_{1}-a_{3}^{2}\right) \rho _{0}^{2}\right) \right] =0.
\label{m3}
\end{equation}%
It is seen from Eq. (\ref{m3}) that complex $K$ appears for $\Omega
\rightarrow +0$ under the condition that
\begin{equation}
\allowbreak 2a_{2}+\left( 4a_{1}-a_{3}^{2}\right) \rho _{0}^{2}>0.
\label{m0}
\end{equation}%
In what follows below, condition (\ref{m0})$\allowbreak $ is referred to as
the \textit{condition of the baseband MI} for Eq. (\ref{5}) under condition (%
\ref{4a}). It is obvious that inequality (\ref{m0}) holds only for $a_{2}>0$%
. In other words, RWs may be excited in Eq. (\ref{5}) by the MI under
condition (\ref{m0}) only in the case of the self-focusing nonlinearity,
with $a_{2}>0$. Specifically, this condition is satisfied for the KE
equation (\ref{6}) in which $a_{2}=2>0$ and $a_{1}=a_{3}^{2}/4$. For any CW
amplitude $\rho _{0}$ that satisfies the baseband-MI condition (\ref{m0}),
Eq. (\ref{m3}) yields $K=2\Omega _{0}+a_{3}\rho _{0}^{2}\pm i\sqrt{\left(
\allowbreak 2a_{2}+\left( 4a_{1}-a_{3}^{2}\right) \rho _{0}^{2}\right) \rho
_{0}^{2}-\Omega ^{2}}$. Therefore, the growth rate (gain) of the baseband MI
is
\begin{equation}
\Gamma (\Omega )=\Omega \sqrt{\left[ \allowbreak 2a_{2}+\left(
4a_{1}-a_{3}^{2}\right) \rho _{0}^{2}\right] \rho _{0}^{2}-\Omega ^{2}}\text{%
, for \ }0\leq \Omega \leq \rho _{0}\sqrt{2a_{2}+\left(
4a_{1}-a_{3}^{2}\right) \rho _{0}^{2}}.  \label{m4}
\end{equation}%
Using Eq. (\ref{m4}), one can obtain the MI map in the plane of $\left(
\Omega ,\rho _{0}\right) $. Figure 1 displays the map associated with Eq. (%
\ref{5}) under condition (\ref{4a}). It is seen from the figure that the
baseband MI occurs not at all values of $\rho _{0}$, and the instability
region expands as parameter $a_{1}$ increases. Note that Fig. 1(d)
corresponds to the KE equation (\ref{6}), and the associated MI boundary
does not depend on $\beta $.

\subsection{ Chirped rogue waves}

\begin{figure}[tbp]
\centerline{\includegraphics[scale=0.85]{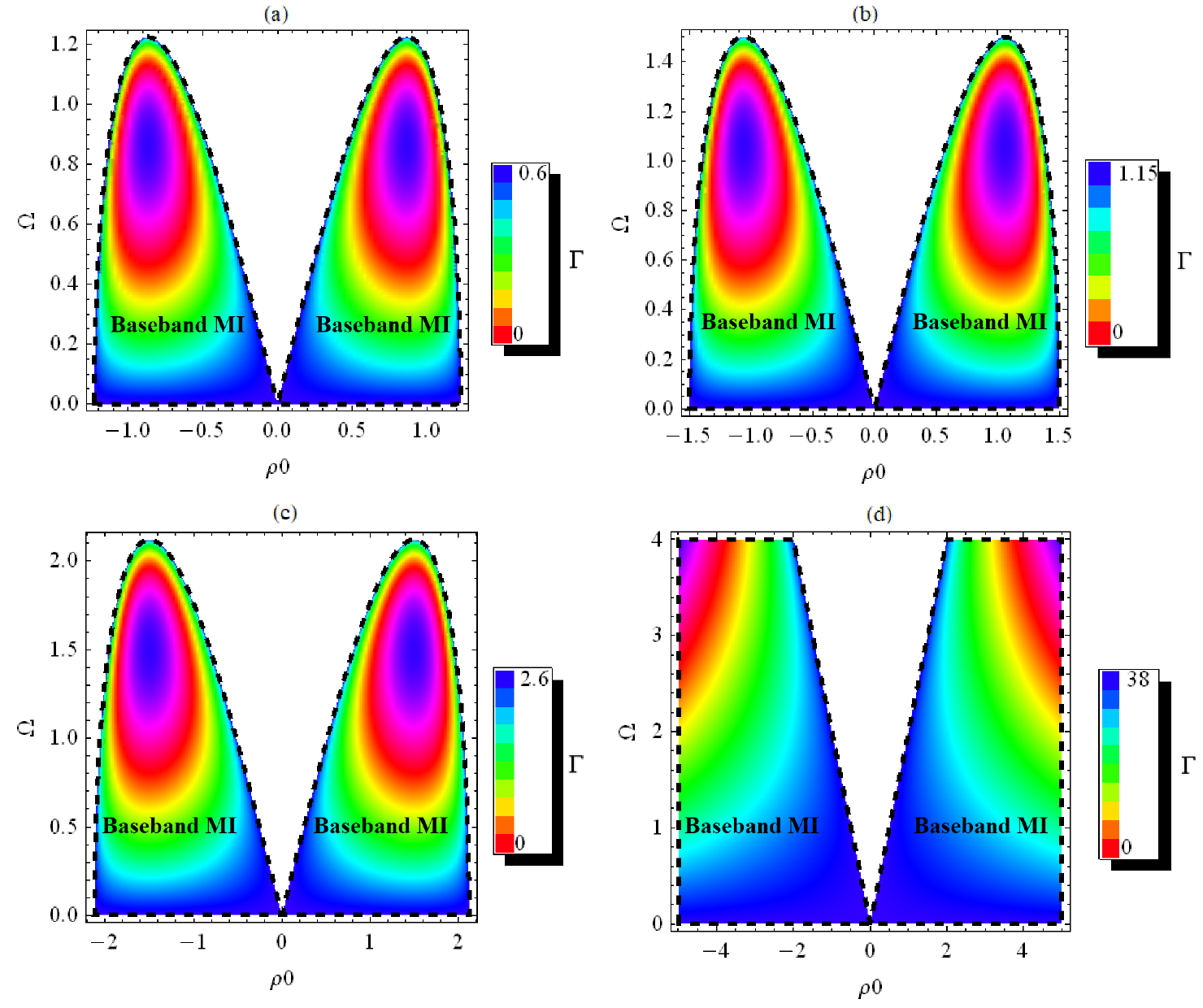}}
\caption{(Color online) MI\ (modulational instability) maps in the plane of $%
\Omega $ and $\protect\rho _{0}$, as produced by Eq. (\protect\ref{m4}) for $%
a_{2}=2,$ $a_{3}=-4\protect\beta $ and four values of $a_{1}$: (a) $a_{1}=-2%
\protect\beta ^{2}$, (b) $a_{1}=0,$ (c) $a_{1}=2\protect\beta ^{2}$, and (d)
$a_{1}=4\protect\beta ^{2}$, with $\protect\beta =1/3$. Dashed lines show MI
boundaries, defined by either $\Omega =\protect\sqrt{\allowbreak 2a_{2}%
\protect\rho _{0}^{2}+\left( 4a_{1}-a_{3}^{2}\right) \protect\rho _{0}^{4}}$
or $\Omega =0$.}
\label{fig1}
\end{figure}

Here, we aim to find exact chirped RW solutions of the cubic GP equation (%
\ref{1}), under the integrability condition (\ref{4e}). We limit ourselves
to the situation when the condition $2a_{2}+\left( 4a_{1}-a_{3}^{2}\right)
\rho _{0}^{2}>0$ of the baseband MI is satisfied. Further, we consider only
the case when the quintic-nonlinearity parameter $a_{1}$ in Eq. (\ref{5}) is
related to the self-frequency-shift one $a_{3}$ by Eq. (\ref{1/4}), and then
we choose the self-frequency-shift coefficient as $a_{3}=-2\sqrt{2a_{2}}%
\beta \neq 0$, so that Eq. (\ref{5}) can be reduced to the integrable KE
equation (\ref{6}), with Fig. 1(d) showing the corresponding MI map. Then,
we exploit the first-order and second-order RW solutions of Eq. (\ref{6})
from \cite{30a,31} to build the analytical first-order and second-order
chirped RW solutions of the cubic GP Eq. (\ref{1}). From the results
obtained in Refs. \cite{30a,31}, the first-order RW solution of Eq. (\ref{5}%
) in the special case with $a_{1}=a_{3}^{2}/4=0$ and $a_{3}=-2\sqrt{2a_{2}}%
\beta \neq 0$ can be written as
\begin{subequations}
\begin{eqnarray}
\Phi (X,T) &=&\frac{4-D_{1}+8ia_{2}T}{D_{1}}\exp \left[ i\left( -\beta \sqrt{%
2a_{2}}X+\frac{a_{2}\left( 2+4\beta ^{2}\right) }{2}T+\beta \sqrt{2a_{2}}%
\int \left\vert \frac{4-D_{1}+8ia_{2}T}{D_{1}}\right\vert ^{2}dX\right) %
\right] ,  \label{17a} \\
D_{1} &=&1+4a_{2}^{2}T^{2}+2a_{2}\left( X+2\sqrt{2a_{2}}\beta T\right) ^{2},
\label{17b}
\end{eqnarray}%
where $a_{2}$ is an arbitrary positive real parameter. Inserting Eq. (\ref%
{17a}) into the ansatz (\ref{2}) and using Eq. (\ref{4e}) yields the
following first-order RW solution of the cubic GP equation (\ref{1}):
\end{subequations}
\begin{subequations}
\begin{eqnarray}
\psi (x,t) &=&\sqrt{\frac{2a(t)}{a_{2}}}\sqrt{\frac{\left( 4-D_{1}\right)
^{2}+64a_{2}^{2}T^{2}}{D_{1}^{2}}}\exp \left[ i\left( \arctan \left[ \frac{%
8a_{2}T}{4-D_{1}}\right] -\frac{1}{4a}\frac{da}{dt}x^{2}-2\beta \sqrt{\frac{2%
}{a_{2}}}a(t)x\right. \right.  \notag \\[0.5cm]
&&\left. \left. +\frac{2\left( 2+4\beta ^{2}\right) }{a_{2}}%
\int_{0}^{t}a^{2}(z)dz-2\theta (x,t)+2\beta \sqrt{\frac{2}{a_{2}}}a(t)\int
\left\vert \frac{4-D_{1}+i8a_{2}T}{D_{1}}\right\vert ^{2}dx\right) \right] ,
\label{18}
\end{eqnarray}%
where $D_{1}=D_{1}(X(x,t),T(t))$ is defined in Eq. (\ref{17b}), with $%
X=X(x,t)$ and $T=T(t)$ given by Eq. (\ref{4f}). The corresponding chirp is%
\begin{equation}
\delta \omega =\frac{1}{2a}\frac{da}{dt}x-2a(t)\sqrt{\frac{2}{a_{2}}}\left(
\beta +\frac{1}{\sqrt{2a_{2}}}\times \left. \frac{\partial }{\partial X}%
\arctan \left[ \frac{8a_{2}T}{4-D_{1}}\right] \right\vert _{X=X(x,t),\text{ }%
T=T(t)}\right) .  \label{20}
\end{equation}

\begin{figure}[tbp]
\centerline{\includegraphics[scale=1.17]{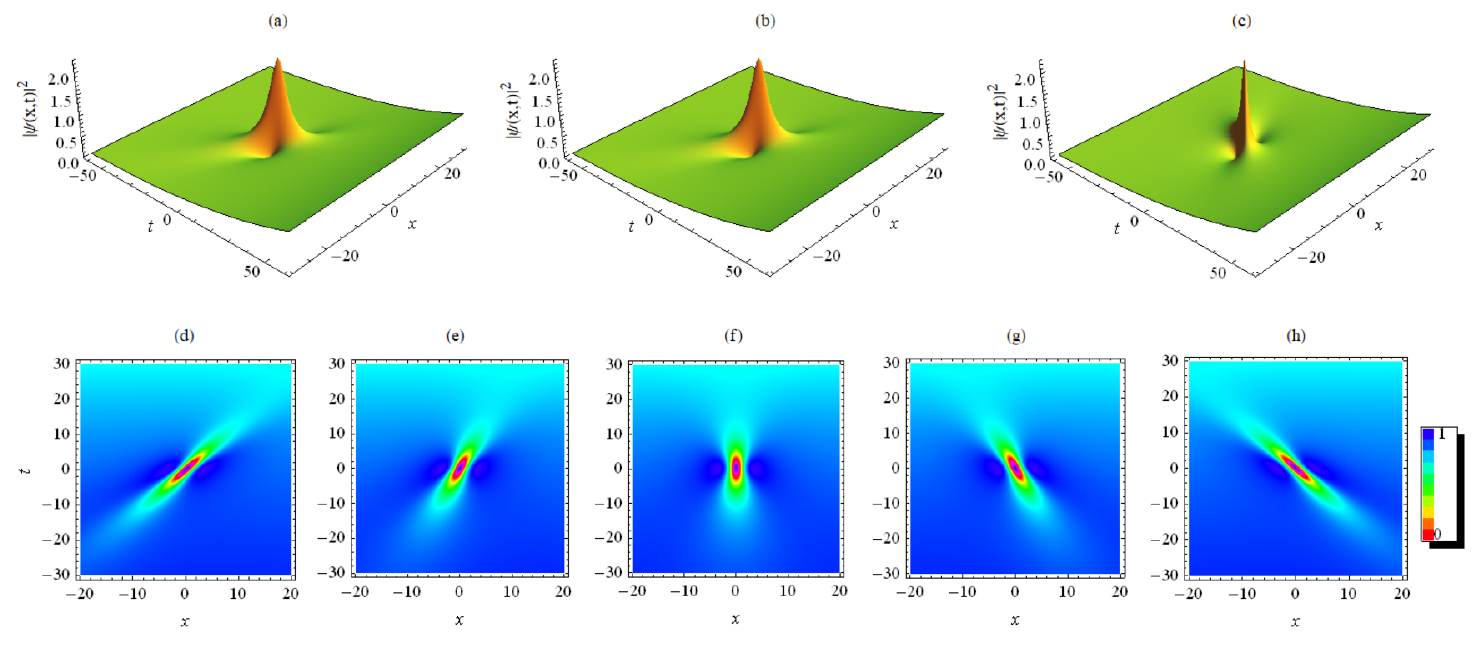}}
\caption{(Color online) Top panels: 3D density plots of the first-order RW
(rogue wave) solution (\protect\ref{18}) for different values of parameter $%
\protect\beta $. (a): $\protect\beta =-1/3$; (b): $\protect\beta =0$; (c): $%
\protect\beta =1/3$. Bottom panels: top views of the density distribution in
the first-order RW solution (\protect\ref{18}) for different values of $%
\protect\beta $. (d): $\protect\beta =-0.7$; (e): $\protect\beta =-1/3$;
(f): $\protect\beta =0$; (g): $\protect\beta =1/3$; (h): $\protect\beta =0.7$%
. Other parameters are $r_{0}=0.25$ and $\protect\lambda =0.02$.}
\label{fig2}
\end{figure}

To derive the second-order RW solution of the cubic GP equation (\ref{1}),
we apply the results from Ref. \cite{30a} for Eq. (\ref{6}), leading to the
following solution of Eq. (\ref{5}):
\end{subequations}
\begin{subequations}
\begin{equation}
\Phi (X,T)=\frac{F_{2}+iG_{2}}{D_{2}}\exp \left[ i\left( -\beta \sqrt{2a_{2}}%
X+\frac{a_{2}\left( 2+4\beta ^{2}\right) }{2}T+\beta \sqrt{2a_{2}}\int
\left\vert \frac{F_{2}+iG_{2}}{D_{2}}\right\vert ^{2}dX\right) \right] ,
\label{21a}
\end{equation}%
where
\begin{eqnarray}
F_{2} &=&45-90a_{2}X^{2}-36a_{2}^{2}X^{4}+8a_{2}^{3}X^{6}-48a_{2}\sqrt{\frac{%
a_{2}}{2}}\left( 15\beta +12\beta a_{2}X^{2}-4\beta a_{2}^{2}X^{4}\right) XT
\notag \\
&&-12a_{2}^{2}\left[ 392+60\beta ^{2}+12a_{2}\left( 5+12\beta ^{2}\right)
X^{2}-4a_{2}^{2}\left( 1+20\beta ^{2}\right) X^{4}\right] T^{2}  \notag \\
&&-128\beta a_{2}^{3}\sqrt{\frac{a_{2}}{2}}\left[ 9\left( 5+4\beta
^{2}\right) -2a_{2}\left( 3+20\beta ^{2}\right) X^{2}\right] XT^{3}
\label{21b} \\
&&-48a_{2}^{4}\left[ 11+120\beta ^{2}+48\beta ^{4}-2a_{2}\left( 1+20\beta
^{2}\right) \left( 1+4\beta ^{2}\right) X^{2}\right] T^{4}  \notag \\
&&+768a_{2}^{5}\beta \sqrt{\frac{a_{2}}{2}}\left( 1+4\beta ^{2}\right)
^{2}XT^{5}+64a_{2}^{6}\beta \left( 1+4\beta ^{2}\right) ^{3}T^{6},  \notag
\end{eqnarray}%
\begin{eqnarray}
G_{2} &=&24a_{2}\left( 15+12a_{2}X^{2}-4a_{2}^{2}X^{4}\right)
T-768a_{2}^{2}\beta \sqrt{\frac{a_{2}}{2}}\left( 2a_{2}X^{3}-3X\right)
T^{2}-192a_{2}^{3}  \notag \\
&&\times \left[ 1-12\beta ^{2}+2a_{2}\left( 12\beta ^{2}+1\right) X^{2}%
\right] T^{3}-3072a_{2}^{4}\beta \sqrt{\frac{a_{2}}{2}}\left( 1+4\beta
^{2}\right) XT^{4}-384a_{2}^{5}\left( 1+4\beta ^{2}\right) ^{2}T^{5},
\label{21c}
\end{eqnarray}%
\begin{eqnarray}
D_{2} &=&9+54a_{2}X^{2}+12a_{2}^{2}X^{4}+8a_{2}^{3}X^{6}+48a_{2}\beta \sqrt{%
\frac{a_{2}}{2}}\left( 4a_{2}^{2}X^{5}+4a_{2}X^{3}+9X\right) T  \notag \\
&&+12a_{2}^{2}\left[ 33+36\beta ^{2}+12a_{2}\left( 2\beta -1\right) \left(
2\beta +1\right) X^{2}+4a_{2}^{2}\left( 1+20\beta ^{2}\right) X^{4}\right]
T^{2}  \notag \\
&&+128a_{2}^{3}\sqrt{\frac{a_{2}}{2}}\left[ 3\beta \left( 4\beta
^{2}-3\right) X+2a_{2}\beta \left( 3+20\beta ^{2}\right) X^{3}\right] T^{3}
\label{21d} \\
&&+48a_{2}^{4}\left[ \left( 4\beta ^{2}-3\right) ^{2}+2a_{2}\left( 1+20\beta
^{2}\right) \left( 1+4\beta ^{2}\right) X^{2}\right] T^{4}  \notag \\
&&+384\beta a_{2}^{5}\sqrt{2a_{2}}\left( 1+4\beta ^{2}\right)
^{2}XT^{5}+64a_{2}^{6}\left( 1+4\beta ^{2}\right) ^{3}T^{6},  \notag
\end{eqnarray}%
where $a_{2}$ is any positive real number, $X=X(x,t)=x/\ell (t)$, and $%
T=T(t) $ is given by Eq. (\ref{4f}), with $\ell (t)$ defined in Eq. (\ref{4a}%
). Combining Eqs. (\ref{2}), (\ref{4e}) and (\ref{21a}) yields the following
second-order RW solution of the cubic GP equation (\ref{1}):
\end{subequations}
\begin{subequations}
\begin{eqnarray}
\psi (x,t) &=&\sqrt{\frac{2a(t)}{a_{2}}}\sqrt{\frac{F_{2}^{2}+G_{2}^{2}}{%
D_{2}^{2}}}\exp \left[ i\left( \arctan \left[ \frac{G_{2}}{F_{2}}\right] -%
\frac{1}{4a}\frac{da}{dt}x^{2}-2\beta \sqrt{\frac{2}{a_{2}}}a(t)x+\frac{%
2\left( 2+4\beta ^{2}\right) }{a_{2}}\right. \right.  \notag \\[0.5cm]
&&\left. \left. \times \int_{0}^{t}a^{2}(z)dz-2\theta (x,t)+2\beta \sqrt{%
\frac{2}{a_{2}}}a(t)\int \left\vert \frac{%
F_{2}(X(x,t),T(t))+iG_{2}(X(x,t),T(t))}{D_{2}(X(x,t),T(t))}\right\vert
^{2}dx\right) \right] .  \label{22a}
\end{eqnarray}%
The corresponding chirp is
\begin{equation}
\delta \omega =\frac{1}{2a}\frac{da}{dt}x-2a(t)\sqrt{\frac{2}{a_{2}}}\left(
\beta +\frac{1}{\sqrt{2a_{2}}}\times \left. \frac{\partial }{\partial X}%
\arctan \left[ \frac{G_{2}}{F_{2}}\right] \right\vert _{X=X(x,t),\text{ }%
T=T(t)}\right) .  \label{22b}
\end{equation}%
In Eqs. (\ref{22a}) and (\ref{22b}), $F_{2},$ $G_{2}$, and $D_{2}$ are given
by Eqs. (\ref{21b}), (\ref{21c}), and (\ref{21d}), respectively, while $%
X=X(x,t)\ $and $T=T(t)$ are defined by Eq. (\ref{4f}).

It is important to note that, if functions $F\left( x,t\right) $ and $G(x,t)$
vanish at point $\left( x,t\right) =\left( x_{0},0\right) $, then expression
\end{subequations}
\begin{equation}
\underset{\left( x,t\right) \rightarrow \left( x_{0},0\right) }{\lim }\left.
\frac{\partial }{\partial X}\arctan \left[ \frac{G}{F}\right] \right\vert
_{X=X(x,t),\text{ }T=T(t)}  \label{lim}
\end{equation}
is undefined at this point. This remark pertains to Eq. (\ref{20}), with $%
F=4-D_{1}$ and $G=8a_{2}T$, and Eq. (\ref{22b}), with $F=F_{2}$ and $G=G_{2}$%
, which give, severally, the chirp of the first- and second-order RWs. It is
demonstrated in the next section that such points $\left( x_{0},0\right) $
are positions of peaks and holes of the chirps.

\begin{figure}[tbp]
\centerline{\includegraphics[scale=1.25]{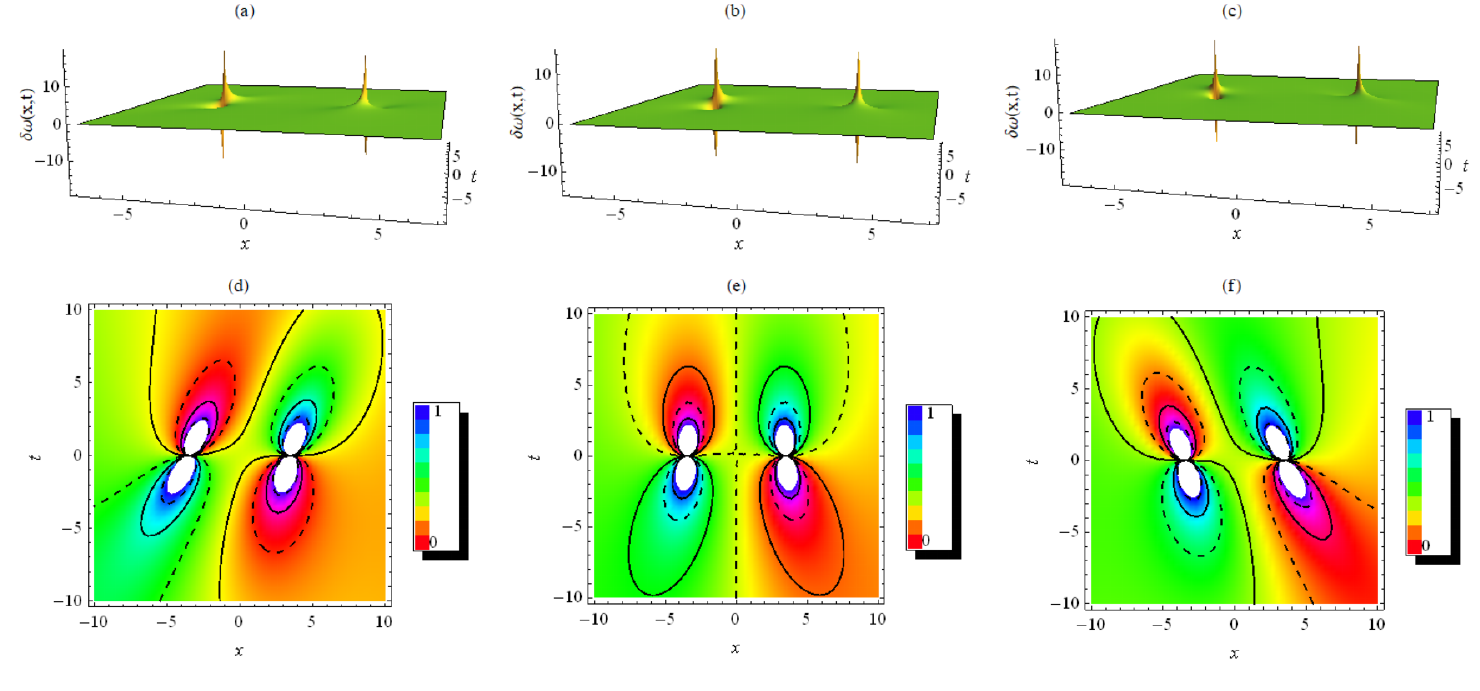}}
\caption{(Color online) 3D (top) and top-view (bottom) distributions of
chirp (\protect\ref{20}) for the first-order RW solution (\protect\ref{18})
with different values of parameter $\protect\beta $. (a,d): $\protect\beta %
=01/3$; (b,e): $\protect\beta =0$; (c,f): $\protect\beta =1/3$. Other
parameters are $r_{0}=0.25$ and $\protect\lambda =0.02$.}
\label{fig3}
\end{figure}

\section{Evolution of chirped rogue waves under the action of the
time-dependent atomic scattering length and parabolic potential}

In this section, we use the above analytical RW solutions of the cubic GP
equation (\ref{1}) to construct chirped RWs in BEC under the consideration.
Throughout this section, without the loss of generality, we take $a_{2}=2$.
From the above first- and second-order RW solutions of Eq. (\ref{1}), we
conclude that (i) the amplitude of the chirped RW is proportional to $\sqrt{%
2a(t)/a_{2}}=\sqrt{a(t)}$, and (ii) chirp $\delta \omega (x,t)$ of the
first- and second-order RW is independent of the phase imprint $\theta (x,t)$%
.

While functions $a(t)$ and $k(t)$ that satisfy the integrability condition (%
\ref{4e}) can be arbitrarily chosen to generate the required chirped RW
solutions of Eq. (\ref{1}), in this work we consider either the trap's
frequency $k(t)$ or the Feshbach-managed nonlinearity coefficient $a(t)$ of
the following forms: (i) $k(t)=-\lambda ^{2}/4$ corresponding to the
time-independent expulsive parabolic potential which was used in the
creation of bright BEC solitons \cite{8}; (ii) $a(t)=r_{0}\left[ 1+m\sin
(\omega t)\right] $ with $0<m<1$ and $r_{0}>0$, corresponding to BEC with a
time-periodic modulation of the \textit{s}-wave scattering length \cite{32};
and (iii) $k(t)=-\left( \mu _{0}^{2}/8\right) \left[ 1-\tanh \left( \mu
_{0}t/2\right) \right] $, where $r_{0}$ is a positive constant \cite{33}.
Then, the integrability condition (\ref{4e}) yields for cases (i), (ii), and
(iii) that one should choose, respectively, $a(t)=r_{0}\exp (\lambda t),$ $%
k(t)=-\left( m\omega ^{2}/4\right) \left[ m+\sin (\omega t)+m\cos
^{2}(\omega t)\right] \left[ 1+m\sin (\omega t)\right] ^{-2}$, and $%
a(t)=r_{0}\left[ 1+\tanh \left( \mu _{0}t/2\right) \right] $ with $r_{0}>0$.
Note that these forms of the time-dependent trap's frequency $k(t)$ or
FR-managed nonlinearity coefficient $a(t)$ are relevant to BEC experiments
\cite{32,33}. Because the amplitude of the above chirped RWs is proportional
to $\sqrt{a(t)}$, parameter $r_{0}$ is referred to as the amplitude. In this
section, we consider each of the three above-mentioned cases separately,
substituting them in the chirped first- and second-order RW solutions (\ref%
{18}) and (\ref{22a}), and the corresponding expressions for the chirp.
Then, we analyze in detail how the chirped RWs get modified by the
time-modulation functions $a(t)$ and $k(t)$.

\begin{figure}[tbp]
\centerline{\includegraphics[scale=0.95]{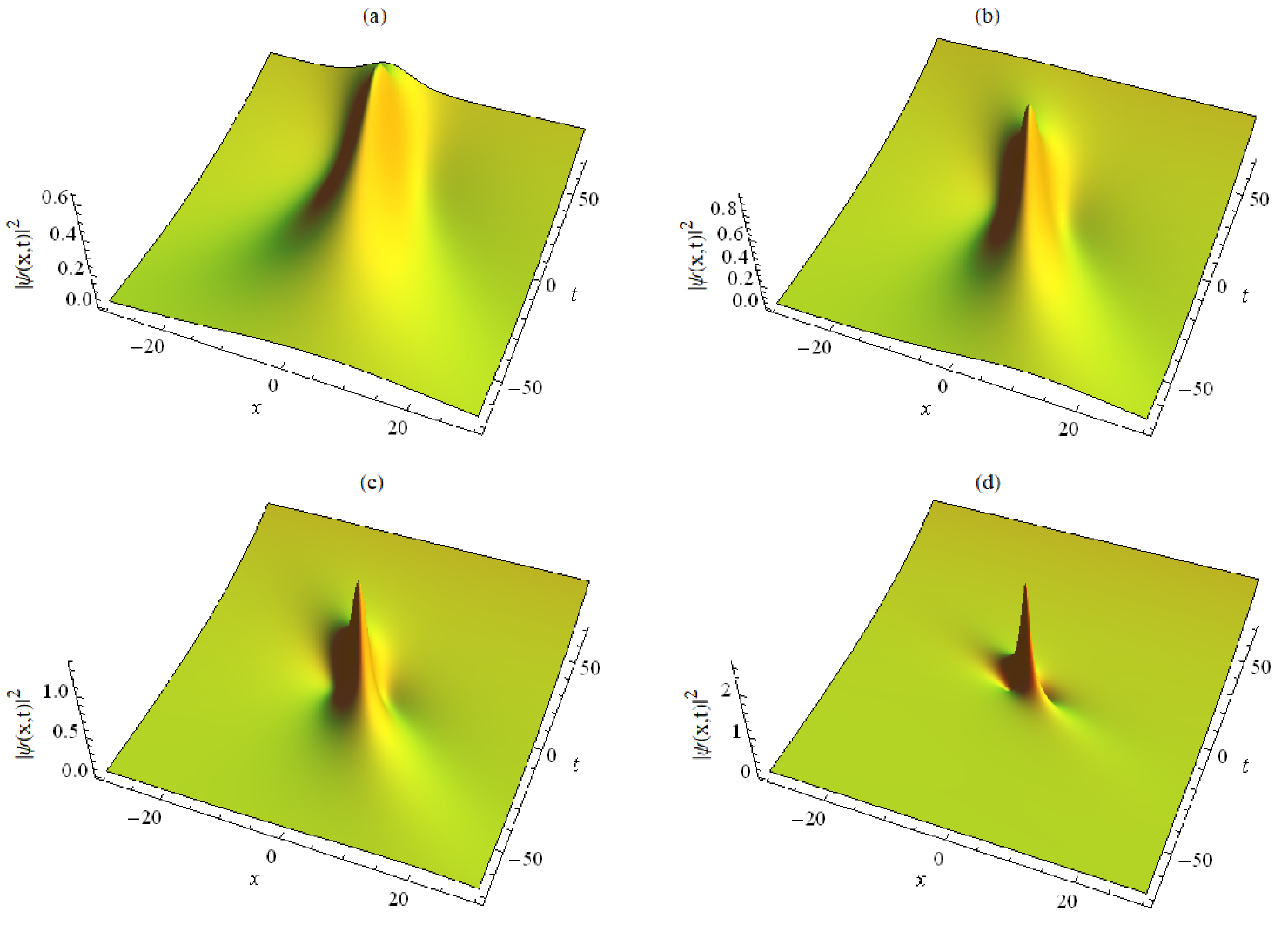}}
\caption{(Color online) The formation of the first-order RWs in BEC with the
nonlinearity coefficient exponentially increasing as per Eq. (\protect\ref%
{exp}) and the time-independent trap's frequency, $k(t)=-\protect\lambda %
^{2}/4$. The plots are generated for $\protect\beta =1/3$, $\protect\lambda %
=0.02$, and different values of the amplitude parameter $r_{0}$ in Eq. (%
\protect\ref{exp}): $r_{0}=0.05$ (a); $r_{0}=0.1$ (b); $r_{0}=0.15$ (c); $%
r_{0}=0.3$ (d).}
\label{fig4}
\end{figure}

\begin{figure}[tbp]
\centerline{\includegraphics[scale=1.17]{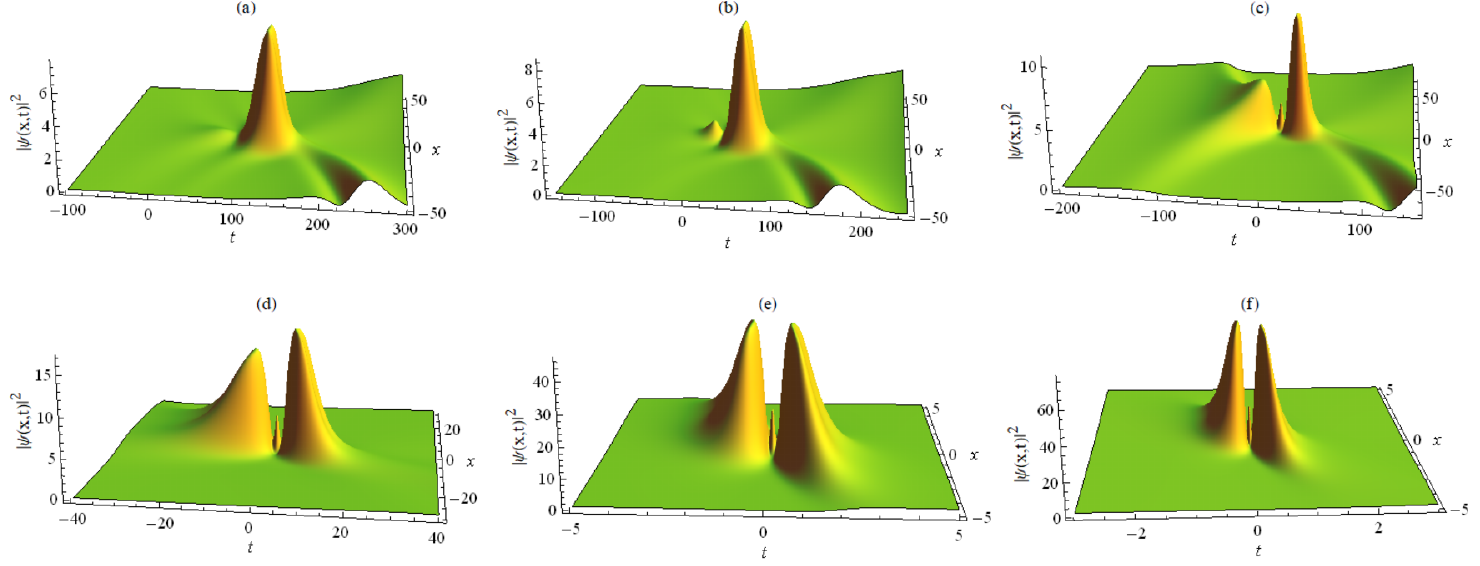}}
\caption{(Color online) Formation of the second-order RWs, in the model with
the same $a(t)$ and $k(t)$ as in Fig. 3, the coefficients being $\protect%
\beta =1/3$ and $\protect\lambda =0.02$. Parameter $r_{0}$ in Eq. (  \protect
\ref{exp}) takes values $r_{0}=0.01$ (a); $r_{0}=0.05$ (b); $r_{0}=0.1$ (c);
$r_{0}=0.2$ (d); $r_{0}=0.6$ (e); $r_{0}=1.0$ (f).}
\label{fig5}
\end{figure}

\subsection{Evolution of chirped rogue waves under the action of the
exponentially varying scattering length and expulsive parabolic potential}

As the first example, we follow the work of Liang \textit{et al}. from Ref.
\cite{7} and consider BEC with an exponentially varying atomic scattering
length in an \textit{expulsive} (\textit{anti-trapping}) time-independent
harmonic potential, described by Eq. (\ref{1}) with $k(t)=-\lambda ^{2}/4$
and
\begin{equation}
a(t)=r_{0}\exp (\lambda t),  \label{exp}
\end{equation}
where $r_{0}>0$. Accordingly, we obtain from Eqs. (\ref{4a}) and (\ref{4f})
that $\ell (t)=r_{0}^{-1}\exp \left( -\lambda t\right) $ and $T(t)=\left(
r_{0}^{2}/2\lambda \right) \left[ \exp \left( 2\lambda t\right) -1\right] $.

In Fig. \ref{fig2}, the top and bottom panels show, respectively, the 3D and top-view
density plots of the first-order RW solution (\ref{18}) for different values
of $\beta $. When $\beta =0$, Eq. (\ref{18}) yields just the standard
Peregrine soliton solution to the standard NLS equation \cite{34}, as seen
in Figs. 2(b) and (f). These solutions feature one hump and two valleys
around the center: the maximum value of the density in the hump is $2.3$,
located at $\left( x,t\right) =(0,0)$, and the minimum value in the valleys
is $0$, located at $\left( x,t\right) =\left. \left( \pm \sqrt{3}/\left(
2r_{0}\right) ,0\right) \right\vert _{r_{0}=0.25}\approx \left( \pm
3.4641,0\right) $. When $\beta \neq 0$, it is seen from Figs. 2(a), (b),
(c), (d), (g), and (h) that the shape of the first-order RW does not change
drastically. However, the three-body interaction term with strength $4\beta
^{2}$ and the delayed nonlinear-response one (the last term in Eq. (\ref{6}%
)) produce an essential skew angle relative to the ridge of the RW in the
anti-clockwise direction for $\beta >0$ (see Figs. 2(g) and (h)), and in the
clockwise direction for $\beta <0$, as seen in Figs. 2(d) and (e). As shown
by panels (d)-(e) and (g)-(h), the skew angle becomes larger with the
increase of $\left\vert \beta \right\vert $. From Figs. 2(a), (b), and (c)
it is seen that the first-order RW for BEC in the expulsive time-independent
parabolic potential are localized in both $x$ and $t$ directions, which
means that the first-order RWs can concentrate the condensate in a small
region. Unlike classical RWs, plots 2(a), 2(b), and 2(c) show that the
nonzero backgrounds of the waves increase with time $t$. This, in turn,
means that the dynamics of RWs in BECs with the exponentially varying atomic
scattering length in the expulsive time-independent parabolic potential is
similar to that in BEC with supply of atoms. This similarity can be
conformed by using transformation $\psi (x,t)=\widetilde{\psi }(x,t)\exp
\left( -\lambda t/2\right) $.

\begin{figure}[tbp]
\centerline{\includegraphics[scale=1.2]{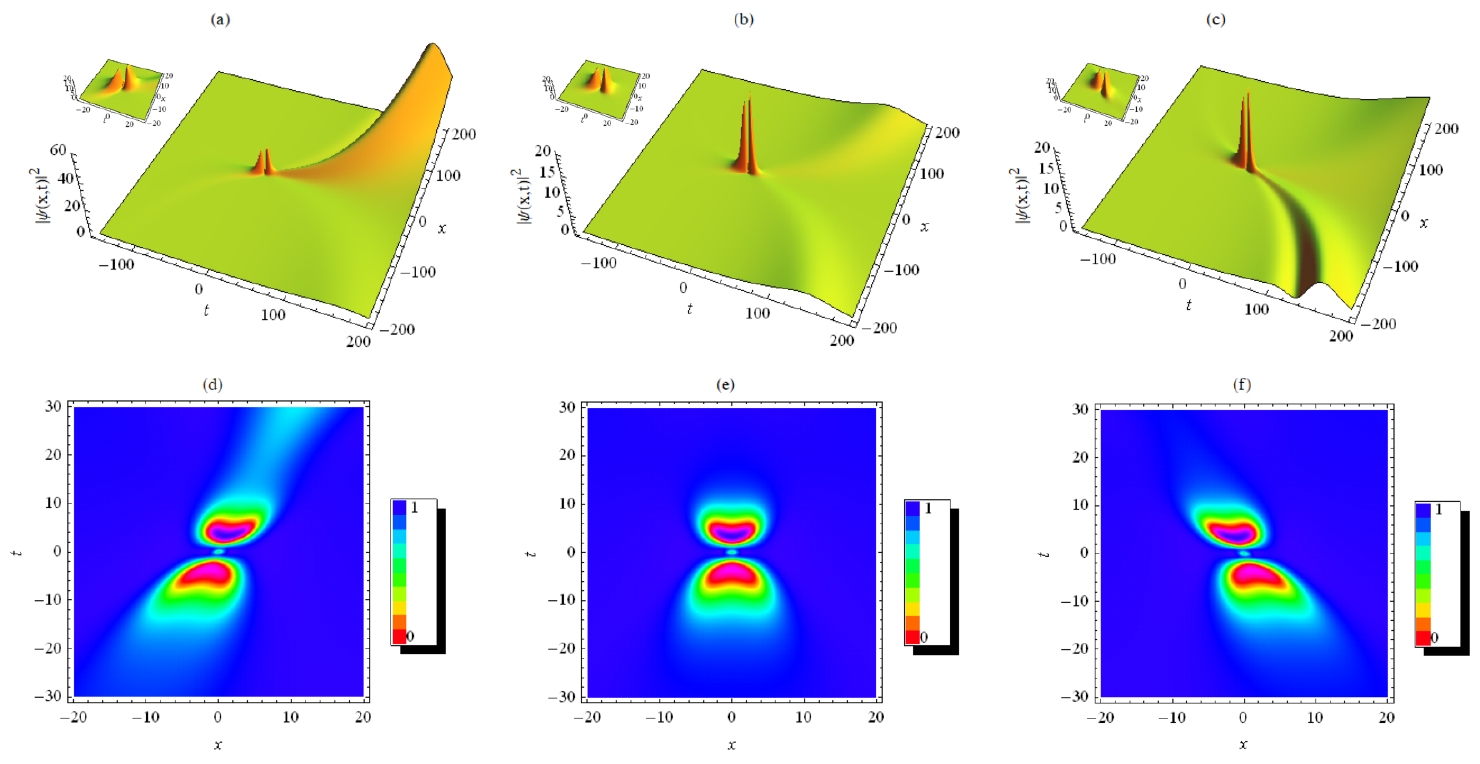}}
\caption{(Color online) Effects of strength $\protect\beta $ of the
three-body interaction on the dynamics of the second-order RW in BEC with
the exponentially varying atomic scattering length (see Eq. (\protect\ref%
{exp}) and expulsive time-independent parabolic potential. The top and
bottom panels are organized as in Fig. \ref{fig2}, but for the solution given by Eq. (%
\protect\ref{22a}), for $\protect\beta =-1/3$ (a,d), $\protect\beta =0$
(b,e), and $\protect\beta =1/3$ (c,f). Parameters $r_{0}$ and $\protect%
\lambda $, as well as the time-dependent ones, $a(t)$ and $k(t)$, are given
in the text.}
\label{fig6}
\end{figure}

Figure 3 displays the typical spatiotemporal distribution of chirp $\delta
\omega (x,t)$, as given by Eq. (\ref{20}), which corresponds to the
first-order RW associated with the analytical solution (\ref{18}). The
figure reveals that the frequency chirp of the first-order RW is localized
in both time and space, exhibiting two dark-bright doubly localized
structures, with the same location as in the two valleys of the
corresponding first-order RW, \textit{viz}. at $\left( x=\pm \sqrt{3}/\left(
2r_{0}\right) ,0\right) $. Top panels (a), (b), and (c) show that the shape
of $\delta \omega (x,t)$ does not change drastically when $\beta \neq 0$
(this is well seen from the comparison of Fig. 3(b) for $\beta =0$ to Figs.
3(a) and 3(c) for $\beta =\pm 1/3$). Nevertheless, the delayed nonlinear
response of the system, with coefficient $\beta $, produces, like in Fig. \ref{fig2},
an essential skew angle relative to the ridge of the frequency chirp in the
anti-clockwise direction for $\beta >0$ (see Fig. 3(c)), and in the
clockwise direction for $\beta <0$, as clearly seen from Fig. 3(a).

\begin{figure}[tbp]
\centerline{\includegraphics[scale=1.25]{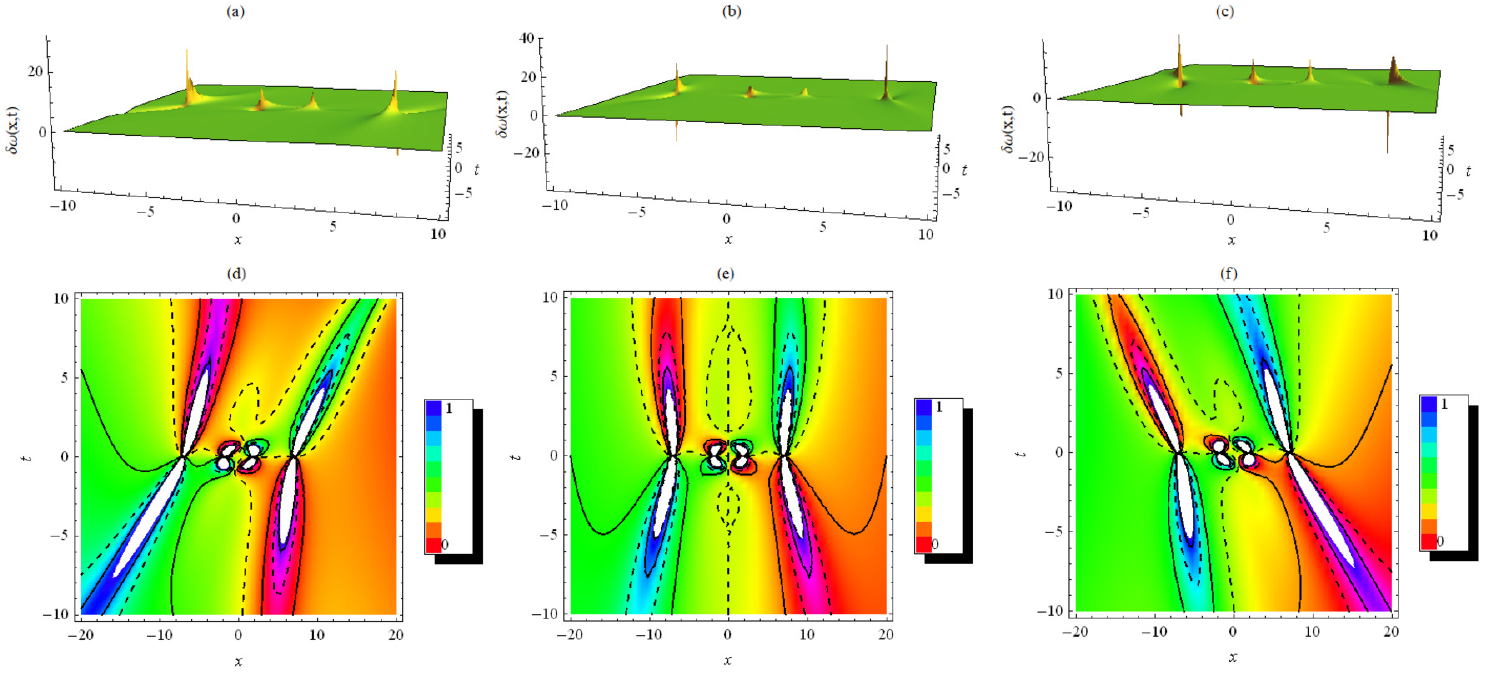}}
\caption{(Color online) 3D (top) and top-view (bottom) distributions of the
chirp for the second-order RW solution (\protect\ref{22a}), given by Eq. (%
\protect\ref{22b}), for $k=-\protect\lambda ^{2}/4$ and $a(t)$ taken as per
Eq. (\protect\ref{exp}), cf. Fig. 3 for the first-order RWs. The results are
plotted for $r_{0}=0.25,$ $\protect\lambda =0.02$, and three values of $%
\protect\beta $, \textit{viz}., $\protect\beta =-0.7$ (a,d), $\protect\beta %
=0$ (b,e), and $\protect\beta =0.7.$(c,f).}
\label{fig7}
\end{figure}

Now, we aim to demonstrate how the first-order RW in BEC with the
exponentially varying atomic scattering length, defined as per Eq. (\ref{exp}%
), in the time-independent expulsive\textit{\ }parabolic potential\ vary
with respect to amplitude parameter $r_{0}$. For this aim, we present in
Fig. 4 the formation of the first-order RWs in the cigar-shaped BEC. By
varying $r_{0}$ from $0.05$ to $0.3$ in the RW solution (\ref{18}), we
visualize the formation and manipulation of first-order RWs. When $r_{0}$
smoothly increases, we observe the formation of crests and troughs, as well
as an increase of the amplitude and decrease of the width of the isolated
wave mode. At $r_{0}=0.3$, a large-amplitude mode is localized in $x$ and $t$%
, which confirms the formation of the first-order RW. Thus, Fig. 4 reveals
that, with the increase of the absolute value of the scattering length
through the amplitude parameter $r_{0}$ in Eq. (\ref{exp}), the peak value
of the first-order RW grows, and its width compresses. Because the quasi-1D
GP equation applies only for low densities, it should be interesting to see
how far one can compress the first-order RW in a real experiment, increasing
$r_{0}$ as in the above consideration.

\begin{figure}[tbp]
\centerline{\includegraphics[scale=1.19]{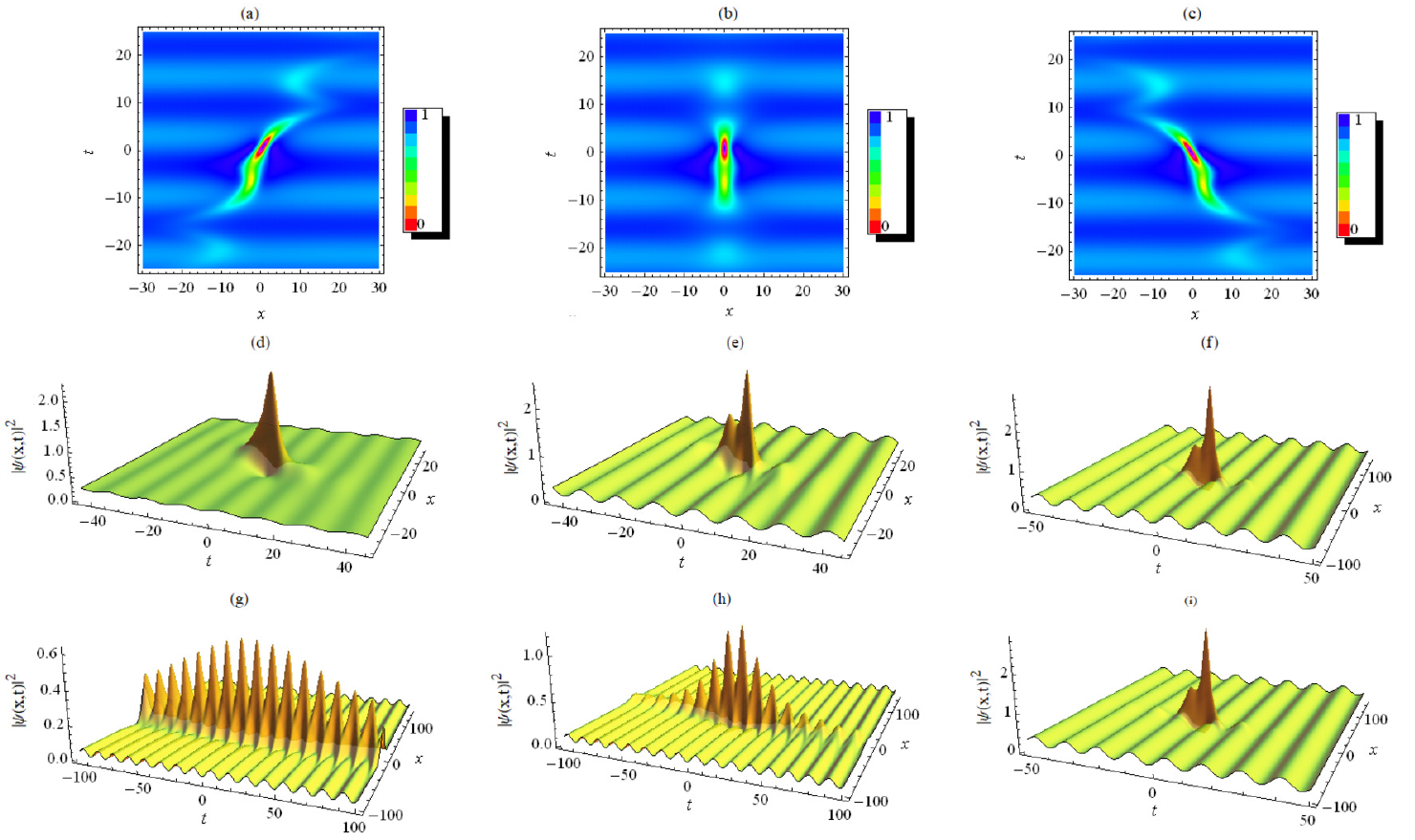}}
\caption{(Color online) Top panels: the density plot for the first-order RW
solution (\protect\ref{18}) with $r_{0}=0.25,$ $m=0.4$, and for different
values of strength $\protect\beta $ of the three-body interaction, \textit{%
viz}., $\protect\beta =-0.7$ (a); $\protect\beta =0$ (b); $\protect\beta %
=0.7 $ (c). Middle panels: 3D density plots for the same RW solution with $%
r_{0}=0.25,$ $\protect\beta =1/3$, and three values of parameter $m$,
\textit{viz}., $m=0.1$ (d); $m=0.4$ (e); $m=0.8$ (f). Bottom panels: density
plots for the same RW solution with $\protect\beta =1/3$, $m=0.4$, and three
values of the amplitude parameter $r_{0}$, \textit{viz}., $r_{0}=0.05$ (g), $%
r_{0}=0.12$ (h); $r_{0}=0.25$ (i). The plots are generated with $\protect%
\omega =0.5$.}
\label{fig8}
\end{figure}

Next, we turn to the consideration of the second-order chirped RW, as per
Eq. (\ref{22a}). It is displayed in Fig. 5, for the time-independent
parabolic potential with the atomic scattering length exponentially varying
according to Eq. (\ref{exp}). As in the case of the first-order RW,
different plots in the figure show that the wave background increases with
time $t$. Depending on the value of $r_{0}$ in Eq. (\ref{exp}), the
fundamental second-order RW can either remain a single RW (for very small
values of $r_{0}$, as seen in Fig. 5(a)), or split into either two
first-order RWs, namely, one small-amplitude and one giant RW, as shown in
Fig. 5(b), or a set of three first-order RWs, including one small-amplitude
and two giant RWs, see Figs. 5(c-f). Thus, Fig. 5 demonstrates the
transformation of the second-order RW with the variation of $r_{0}$. At $%
r_{0}=0.01$, it transforms into the first-order-RW-like structure in Fig.
5(a). When $r_{0}$ increases to $0.05$, a small-amplitude RW (the \textit{%
second RW}) emerges, coexisting with a giant RW (the \textit{first RW}), as
seen in Fig. 5(b). When $r_{0}$ increases to some $\widetilde{r}_{0}$, a
\textit{third RW} emerges, coexisting with the first and the second ones, as
observed in Fig. 5(c) for $r_{0}=0.1>\widetilde{r}_{0}.$ At all values $%
r_{0}>\widetilde{r}_{0}$, the three RWs coexist, their amplitudes increasing
with the growth of $r_{0}$. This is observed in Figs. 5(d), (e), and (f) for
$r_{0}=0.2$, $r_{0}=0.6$, and $r_{0}=1.0$, respectively. As seen in Fig.
5(f), the two giant RWs have almost equal amplitudes for large values of $%
r_{0}$ (for values slightly exceeding $\widetilde{r}_{0}$, the amplitude of
the first RW is larger than that of the third one, as seen in Fig. 5 (d)).
For $r_{0}=0.2$ in Fig. 5(d), the first, the second, and third RWs are
located, approximately, at $\left( x,y\right) \approx \left( 0,5.47\right) ,$
$\left( 0,-6.08\right) $, and $\left( 0,-1.16\right) $, respectively, their
amplitudes being, respectively, $16.79$, $1.34$, and $13.29$. For $r_{0}=1$
in Fig. 5(f), our calculations show that the maximum amplitude of the first,
second, and third RWs are, respectively, $75.80$, $75.102$, and $25$. These
maxima are attained, severally, at $\left( x,y\right) \approx \left(
0,0.23186\right) $, $\left( 0,-0.23289\right) $, and $\left(
0,0.00001\right) $. Simultaneously, the width of the modes decreases with
the increase of $r_{0}$. Therefore, for BEC with atomic scattering length
exponentially varying as per Eq. (\ref{exp}) and the expulsive
time-independent parabolic potential, Eq. (\ref{22a}) can be used to predict
the compression of second-order RWs.

\begin{figure}[tbp]
\centerline{\includegraphics[scale=1.17]{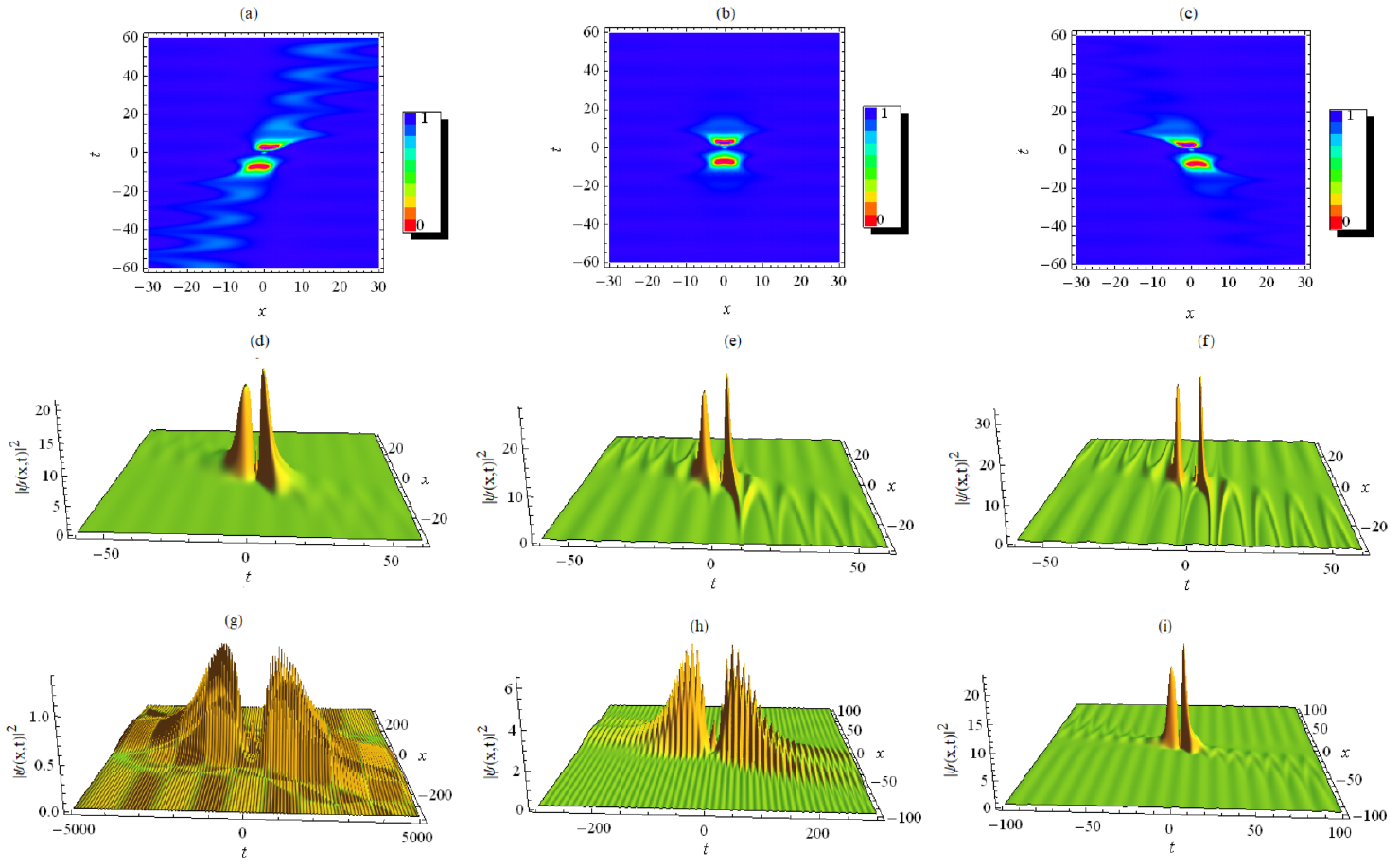}}
\caption{(Color online) Top panels: density plots of the second-order RW
solution (\protect\ref{22a}) with $r_{0}=0.25,$ $m=0.4$, and for different
values of strength $\protect\beta $ of the three-body interaction, \textit{%
viz}., $\protect\beta =-1/3$ (a); $\protect\beta =0$ (b); $\protect\beta %
=1/3 $ (c). Middle panels: 3D density plots of the same RW solution with $%
r_{0}=0.25,$ $\protect\beta =1/3$, and three values of parameter $m$,
\textit{viz}., $m=0.1$ (d); $m=0.5$ (e); $m=0.9$ (f). Bottom panels: 3D
density plots for the same RW solution for $\protect\beta =1/3$, $m=0.2$,
and three values of the amplitude parameter $r_{0}$, \textit{viz}., $%
r_{0}=0.01$ (g); $r_{0}=0.05$ (h); $r_{0}=0.25$ (i).}
\label{fig9}
\end{figure}

As in the case of the first-order RW, parameter $\beta $ of the three-body
interaction does not dramatically affect the shape of the second-order RW,
but produces an essential skew angle relative to the RW's ridge in the
anti-clockwise direction for $\beta >0$, and in the clockwise direction for $%
\beta <0$, cf. Figs. 2 and 3. In Fig. 6 we plot the second-order RW solution
(\ref{22a}) for different values of $\beta $, with $r_{0}=0.25$ and $\lambda
=0.02$. Note that each RW in Figs. 6(a), (b), and (c) features three humps,
although only the two giant ones are clearly visible. The second-order RW
given by Eq. (\ref{22a}) has four valleys at positions $\left( x,y\right)
=\left( x_{0},0\right) $, where $x_{0}=x_{0}(r_{0})$ are real solutions of a
cubic equation (with respect to $\left( 2r_{0}x\right) ^{2}$), $%
45a_{2}^{3}-90a_{2}^{2}\left( 2r_{0}x\right) ^{2}-36a_{2}\left(
2r_{0}x\right) ^{4}+8\left( 2r_{0}x\right) ^{6}=0$. The minimum value of the
solution in the four valleys is $0$. For $r_{0}=0.25$ and $a_{2}=2$, the
four valleys are located around the center at positions $\left( x,y\right)
\approx \left( -7.028,0\right) $, $\left( -1.859,0\right) $, $\left(
1.859,0\right) $, and $\left( 7.028,0\right) $.

The frequency chirp $\delta \omega (x,t)$ given by Eq. (\ref{22b}), which
corresponds to the second-order RW solution (\ref{22a}), is displayed in
Fig. 7. This figure reveals that the chirp is localized both in time and
space. Furthermore, it exhibits four dark-bright doubly localized structures
around $x=0$, located at the same position as the four valleys of the
corresponding second-order RW (\ref{22a}), that is, at $\left( x,y\right)
=\left( x_{0}(r_{0}),0\right) $, $x_{0}(r_{0})$ being real solutions of the
cubic equation $45a_{2}^{3}-90a_{2}^{2}\left( 2r_{0}x\right)
^{2}-36a_{2}\left( 2r_{0}x\right) ^{4}+8\left( 2r_{0}x\right) ^{6}=0$. As in
the case of the first-order RW, the shape of the frequency chirp $\delta
\omega (x,t)$ does not dramatically change at $\beta \neq 0$, as seen from
comparison of Fig. 7(b), generated for $\beta =0$, with Figs. 7(a) and (c)
obtained for $\beta =\pm 0.7$. It is seen in the bottom panels of Fig. 7
that parameter $\beta $ determining the delayed nonlinear response produces,
as above, an essential skew angle relative to the ridge of the frequency
chirp in the anti-clockwise direction for $\beta >0$, see Fig. 7(c), and in
the clockwise direction for $\beta <0$, in Fig. 7(a).

The peaks/holes of the chirp associated to the second-order RW solution (\ref%
{22a}) are located at the same position $\left( x_{0},0\right) $ where the
corresponding valleys are placed, see the note attached to Eq. (\ref{lim}).
For a better presentation of this feature, in Figs. 7(a), (b), and (c) we
show the spatiotemporal evolution of the corresponding chirp.

\subsection{Evolution of chirped rogues waves under the action of
time-periodic modulation of the scattering length}

Now, we investigate, as another case of general interest, chirped RWs in the
BEC model with a temporally periodic variation of the \textit{s}-wave
scattering length \cite{32}, with
\begin{equation}
a(t)=r_{0}\left[ 1+m\sin \left( \omega t\right) \right] ,  \label{sin}
\end{equation}
where we set $0<m<1$ and $r_{0}>0$. The corresponding potential strength $%
k(t)$, that satisfies the integrability condition (\ref{4e}), is taken as
\begin{equation}
k(t)=-\frac{m\omega ^{2}\left[ m+\sin \left( \omega t\right) +m\cos
^{2}\left( \omega t\right) \right] }{4\left[ 1+m\sin \left( \omega t\right) %
\right] ^{2}}.  \label{ksin}
\end{equation}
Because the amplitude of the chirped RW is proportional to $\sqrt{2a(t)/a_{2}%
}$, we conclude that the wave's amplitude will increase with both $r_{0}$
and $m$.

\begin{figure}[tbp]
\centerline{\includegraphics[scale=1.25]{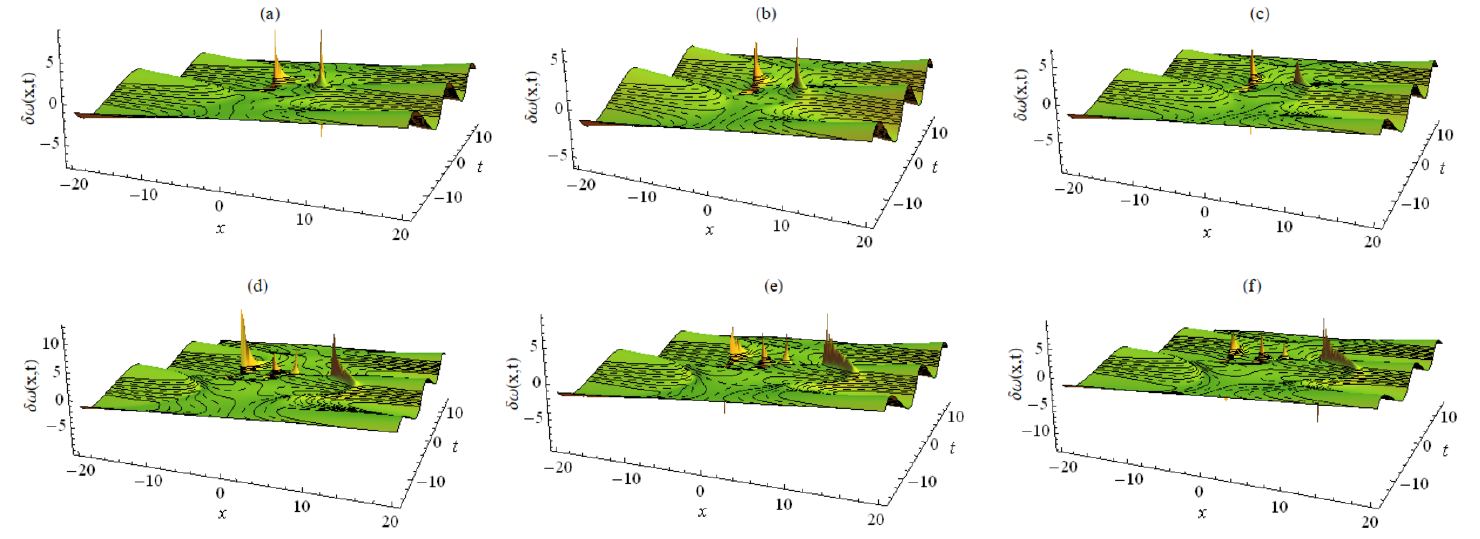}}
\caption{(Color online) Top panels: the distribution of the frequency chirp (%
\protect\ref{20}) corresponding to the first-order RW solution (\protect\ref%
{18}). Bottom panels: the distribution of the frequency chirp (\protect\ref%
{22b}) for the second-order RW solution (\protect\ref{22a}). The plots are
generated with $r_{0}=0.25,$ $m=0.4$, $\protect\omega =0.5,$ and three
different values of strength $\protect\beta $ of the delayed nonlinear
response, \textit{viz}., $\protect\beta =-1/3$ (a,d); $\protect\beta =0$
(b,e); $\protect\beta =1/3\ $(c,f).}
\label{fig10}
\end{figure}

For parameters $a(t)$ and $k(t)$ with the time dependence defined as per
Eqs. (\ref{sin}) and (\ref{ksin}), we display, in Figs. 8, 9, and 10, the
first-order RWs, the second-order RWs, and the corresponding chirp,
respectively. These figures show that, due to the periodic modulation, the
first- and second-order RWs, as well as the corresponding chirp propagate on
top of the modulated CW background. For $\beta =-1/3$, the first- and
second-order RWs are shown, respectively, in Figs. 8(a) and 9(a). For $\beta
=0$, RWs of the same types are displayed in Figs. 8(b) and 9(b),
respectively. Finally, for $\beta =1/3$, the solutions are presented in
Figs. 8(c) and 9(c). From these figures, one can conclude that the
three-body interaction with strength $\beta $ produces, as above, an
essential skew angle relative to the ridge of the RW, in the clockwise
direction for $\beta <0$, and in the anti-clockwise direction for $\beta >0$%
. For different values of parameters $m$ and $r_{0}$ in Eq. (\ref{sin}) $%
a(t) $, we plot, severally, in the middle and bottom panels of Figs. 8 and 9
the first-order RW solution given by Eq. (\ref{18}), and the second-order RW
solution (\ref{22a}). These plots demonstrate how parameters $m$ and $r_{0}$
affect the amplitude and structure of the first- and second-order RWs. In
particular, the modes' amplitudes increase as $m$ and $r_{0}$ grow. It is
seen from the plots in the middle panels that the best structure of the
first- and second-order RWs is obtained for small values of $m$, see Figs.
8(d) and 9(d). Further, the bottom panels in Figs. 8 and 9 reveal that the
best structure of the RWs is attained for higher values of $r_{0}$, see Fig.
8(i) and 9(i). Therefore, parameters $m$ and $r_{0}$ in Eq. (\ref{sin}) have
the same effect on the waves' amplitudes and opposite effect on their
structure, in the sense that the best structure is achieved either by
decreasing $m$ or increasing $r_{0}$. Figure 8(g) also demonstrates that,
for small values of $r_{0}$, the first-order RW behaves like a breather
soliton propagating on top of a modulated CW background.

The top and bottom panels of Fig. 10 showing the distribution of the
frequency chirp which corresponds, respectively, to the first- and
second-order RWs, it is clearly seen that, under the action of the
time-periodic modulation of the \textit{s}-wave scattering length, the chirp
is localized in time and space on top of the CW background. Further, the top
and bottom panels in the figure reveal that the chirp corresponding,
respectively, to the first- and second-order RWs exhibit two- and four-peak
dark-bright doubly localized structures, located (as above) at the same
positions where valleys of the corresponding RWs are found. The plots
related to different signs ($-$, $0$, or $+$) of coefficient $\beta $ of the
delayed nonlinear response in the system do not strongly affect the shape of
the RWs. Nevertheless, also similar to what was observed in several patterns
displayed above, $\beta $ determines an essential skew angle relative to the
ridge of the chirp in the counter-clockwise direction for $\beta >0$, and in
the clockwise direction for $\beta <0$ (in the present case, this feature,
typical for all the RW patterns considered in this work, is not shown in
detail).

\begin{figure}[tbp]
\centerline{\includegraphics[scale=1.21]{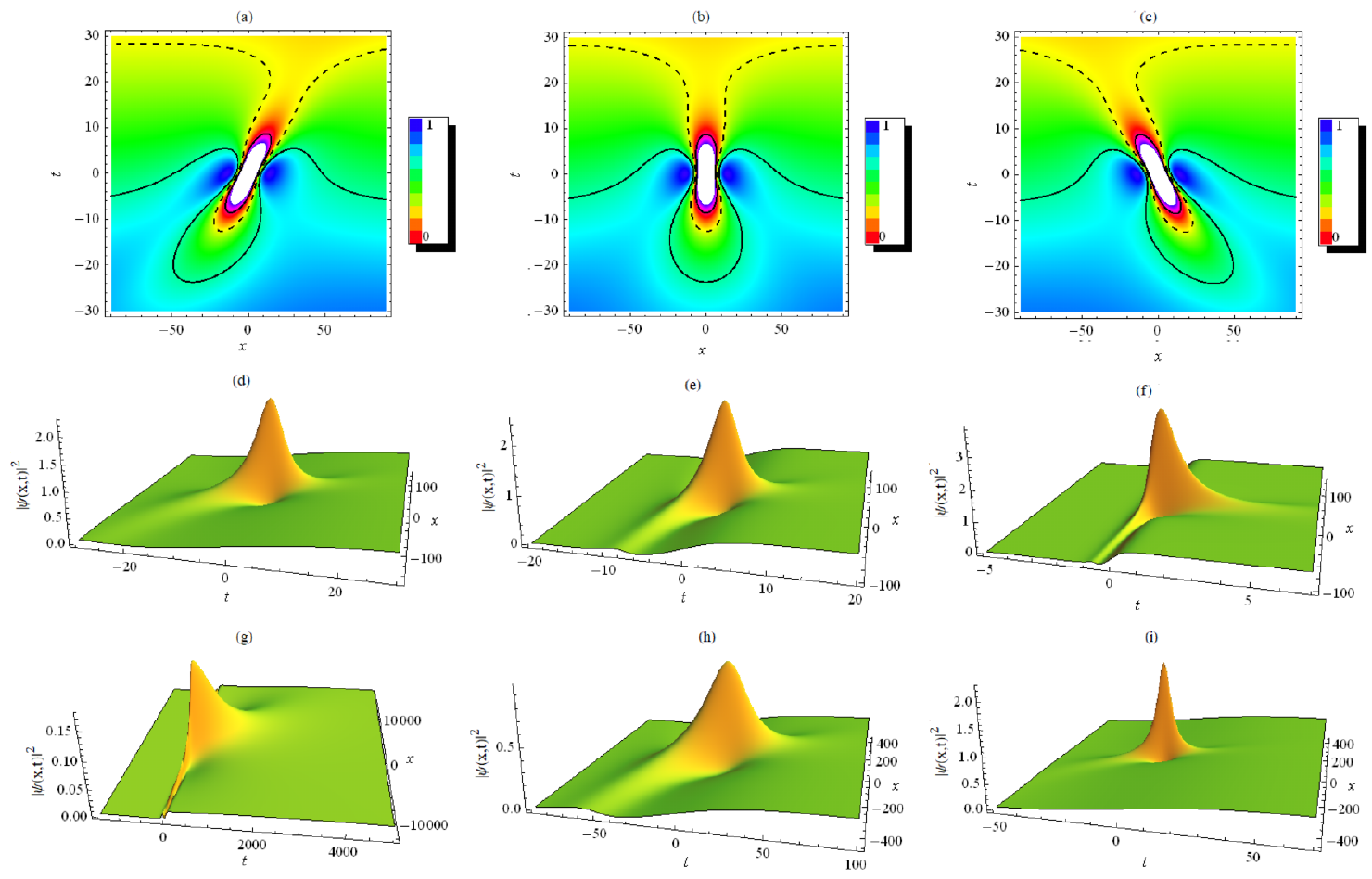}}
\caption{(Color online) Density plots for the first-order RW solution (%
\protect\ref{18}) of Eq. (\protect\ref{1}) with the time-modulation format
given by Eqs. (\protect\ref{step}) and (\protect\ref{astep}). Top panels:
the plots for $r_{0}=0.25,$ $\protect\mu _{0}=0.05$, and different values of
strength $\protect\beta $ of the three-body interaction, \textit{viz}., $%
\protect\beta =-1/3$ (a); $\protect\beta =0$ (b); $\protect\beta =1/3$ (c).
Middle panels: the plots for $r_{0}=0.25,$ $\protect\beta =-1/3$, and three
values of $\protect\mu _{0}$, \textit{viz}., $\protect\mu _{0}=0.1$ (d); $%
\protect\mu _{0}=0.4$ (e); $\protect\mu _{0}=5$ (f). Bottom panels: the
plots for $\protect\beta =-1/3$, $\protect\mu _{0}=0.05$, and three values
of the $r_{0}$, \textit{viz}., $r_{0}=0.05$ (g), $r_{0}=0.1$ (h); $%
r_{0}=0.25 $ (i).}
\label{fig11}
\end{figure}

\subsection{Chirped rogue waves in under the action of the scattering length
and expulsive parabolic potential subjected to the stepwise temporal
modulation}

As the third example, we consider chirped RWs controlled by a time-dependent
scattering length combined with the expulsive parabolic potential whose
strength, $k(t)$, vanishes at $t\rightarrow \infty $. Following Ref. \cite%
{33,34}, we choose a stepwise modulation profile satisfying this condition:%
\begin{equation}
k(t)=-\frac{\mu _{0}^{2}}{8}\left[ 1-\tanh \left( \frac{\mu _{0}}{2}t\right) %
\right] ,  \label{step}
\end{equation}%
with $\mu _{0}>0$. Substituting this in integrability condition (\ref{4e}),
we find the respective time-modulation form of the nonlinearity coefficient,
\begin{equation}
a(t)=r_{0}\left[ 1+\tanh \left( \frac{\mu _{0}}{2}t\right) \right] ,
\label{astep}
\end{equation}%
where $r_{0}$ is an arbitrary positive constant. Because the amplitude of
the first- and second-order chirped RWs, given by Eqs. (\ref{18}) and (\ref%
{22a}), is proportional to $\sqrt{2a(t)/a_{2}}=\sqrt{r_{0}\left[ 1+\tanh
\left( \mu _{0}t/2\right) \right] }$, we conclude that the waves' amplitudes
increase with the growth of $r_{0}$ and $\mu _{0}$.

\begin{figure}[tbp]
\centerline{\includegraphics[scale=1.2]{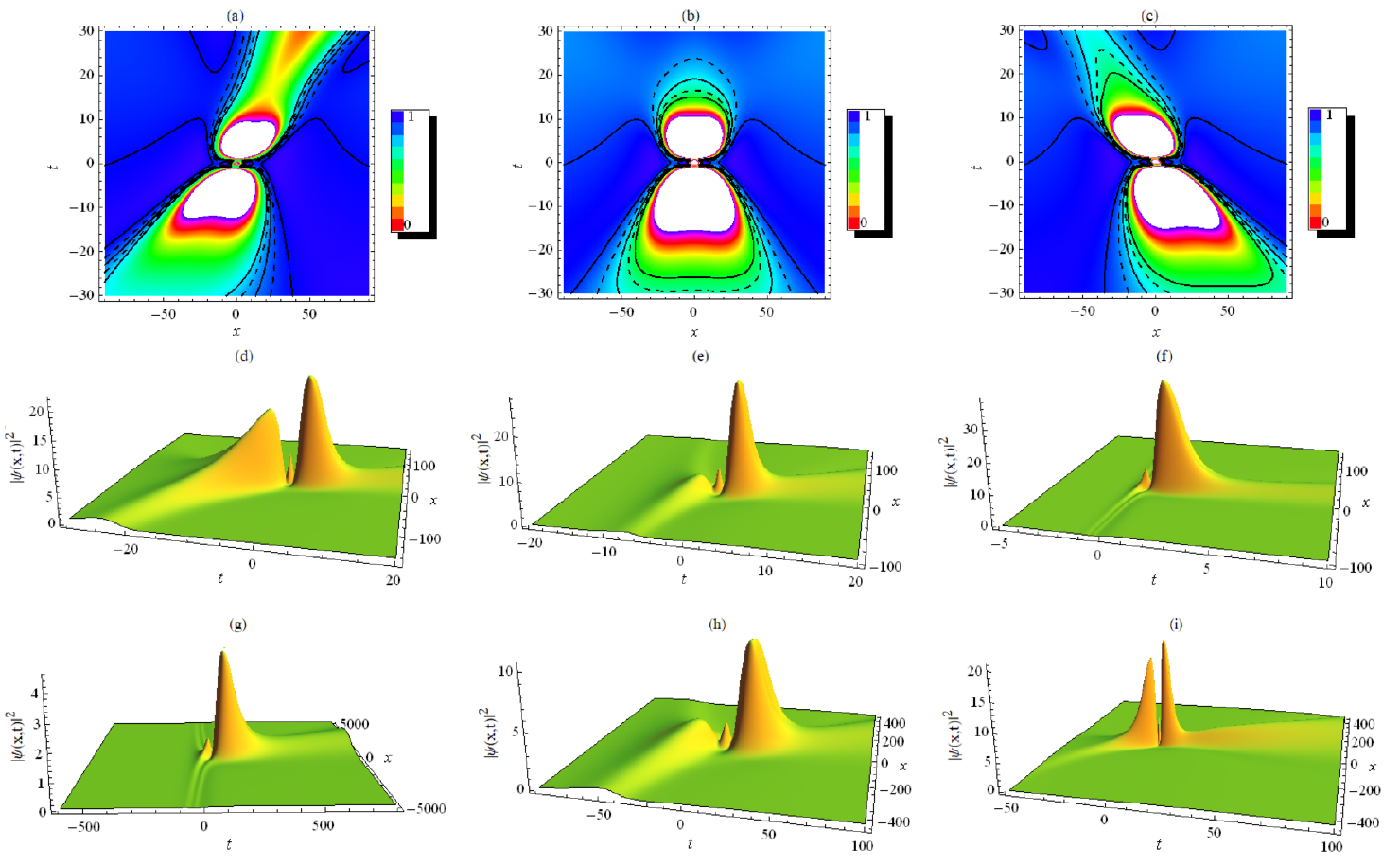}}
\caption{(Color online) Density plots for the second-order RW solution (%
\protect\ref{22a}) with the time-modulation format, defined by Eqs. (
\protect\ref{step}) and (\protect\ref{astep}). Top panels: the plots for $%
r_{0}=0.25, $ $\protect\mu _{0}=0.05$, and different values of strength $%
\protect\beta $ of the three-body interaction, \textit{viz}., $\protect\beta %
=-1/3$ (a); $\protect\beta =0$ (b); $\protect\beta =1/3$ (c). Middle panels:
the plots for $r_{0}=0.25,$ $\protect\beta =-1/3$, and different values of $%
\protect\mu _{0}$, \textit{viz}., $\protect\mu _{0}=0.1$ (d); $\protect\mu %
_{0}=0.4$ (e); $\protect\mu _{0}=5$ (f). Bottom panels: the plots for $%
\protect\beta =-1/3$, $\protect\mu _{0}=0.05$, and three values of $r_{0}$,
\textit{viz}., $r_{0}=0.03$ (g), $r_{0}=0.1$ (h); $r_{0}=0.25$ (i).}
\label{fig12}
\end{figure}

To present effects of modulation parameters $r_{0}$ and $\mu _{0}$ in Eqs. (%
\ref{step}) and (\ref{astep}) on the chirped first- and second-order RWs, in
Figs. 11, 12, and 13 we display the first-order RW given by Eq. (\ref{18}),
the second-order RW given by Eq. (\ref{22a}), and the first- and
second-order chirped RWs produced by Eqs. (\ref{20}) and (\ref{22b}),
respectively. In Figs. 11 and 12, the RWs (\ref{18}) and (\ref{22a}) with $%
\beta =0$ reduce to the standard Peregrine soliton and the standard
second-order RW solution of the integrable NLS equation, see Figs. 11(a) and
12(a). Further, the middle and bottom panels of Figs. 11 and 12 reveal that,
under the action of the temporal modulation defined by Eqs. (\ref{step}) and
(\ref{astep}), the first- and second-order RWs given by Eqs. (\ref{18}) and (%
\ref{22a}) propagate on top of a kink-shaped background. It is also seen in
the top panels of Figs. 11 and 12 that quintic coefficient $\beta $ from Eq.
(\ref{6}) produces, similar to what was seen in the above solutions, an
essential skew angle relative to the ridge of the RW in the clockwise
direction for $\beta >0$ (Figs. 11(a) and 12 (a)), and in the anti-clockwise
direction for $\beta >0$ (Figs. 11(c) and 12(c)). As $\left\vert \beta
\right\vert $ gets larger, the skew angle becomes larger too (not shown here
in detail). It is seen in the middle and bottom panels of Fig. 12 that the
first-order RW, given by Eq. (\ref{18}), with the modulation format (\ref%
{step}) and (\ref{astep}), is composed of one hump and two valleys located
around the center: the amplitude of the hump is $18r_{0}/a_{2}=9r_{0}$,
attained at $(x,y)=\left( 0,0\right) $, while the minima in the valleys is $%
0 $, located at $\left( x,y\right) =\left( \pm \sqrt{3}/\left( 2r_{0}\right)
,0\right) $. Obviously, $\left( \pm \sqrt{3}/\left( 2r_{0}\right) ,0\right)
\rightarrow \left( 0,0\right) $ as $r_{0}\rightarrow +\infty $, meaning
that, for large $r_{0}$, the first-order RW contains a single hump, as seen
in Fig. 11(f), plotted for $r_{0}=5$. Also, $\left( \pm \sqrt{3}/\left(
2r_{0}\right) ,0\right) \rightarrow \left( \pm \infty ,0\right) $ at $%
r_{0}\rightarrow +0$, hence the two valleys of the first-order RW escape to
infinity; in this case, the first-order RW also contains one hump, as seen
in Fig. 11(g), plotted for $r_{0}=0.01$. The middle and bottom panels of
Fig. 11, obtained for different values of,\ respectively, parameters $\mu
_{0}$ and $r_{0}$, show that the amplitude of the first-order RW, given by
Eq. (\ref{18}), increases with each of these two parameters. These panels of
Fig. 11 also show that the mode's width decreases with the increase of $\mu
_{0}$ and $r_{0}$. Therefore, these parameters can be used to control both
the amplitude and width of the first-order RW corresponding to the
modulation format based on Eqs. (\ref{step}) and (\ref{astep}). The middle
and bottom panels of Fig. 12 show that the best structure of the first-order
RW, composed of the single hump and two valleys, is obtained with small
values of $\mu _{0}$ (see Fig. 11(d)) and large values of $r_{0}$ (see Fig.
11 (f)).

\begin{figure}[tbp]
\centerline{\includegraphics[scale=1.15]{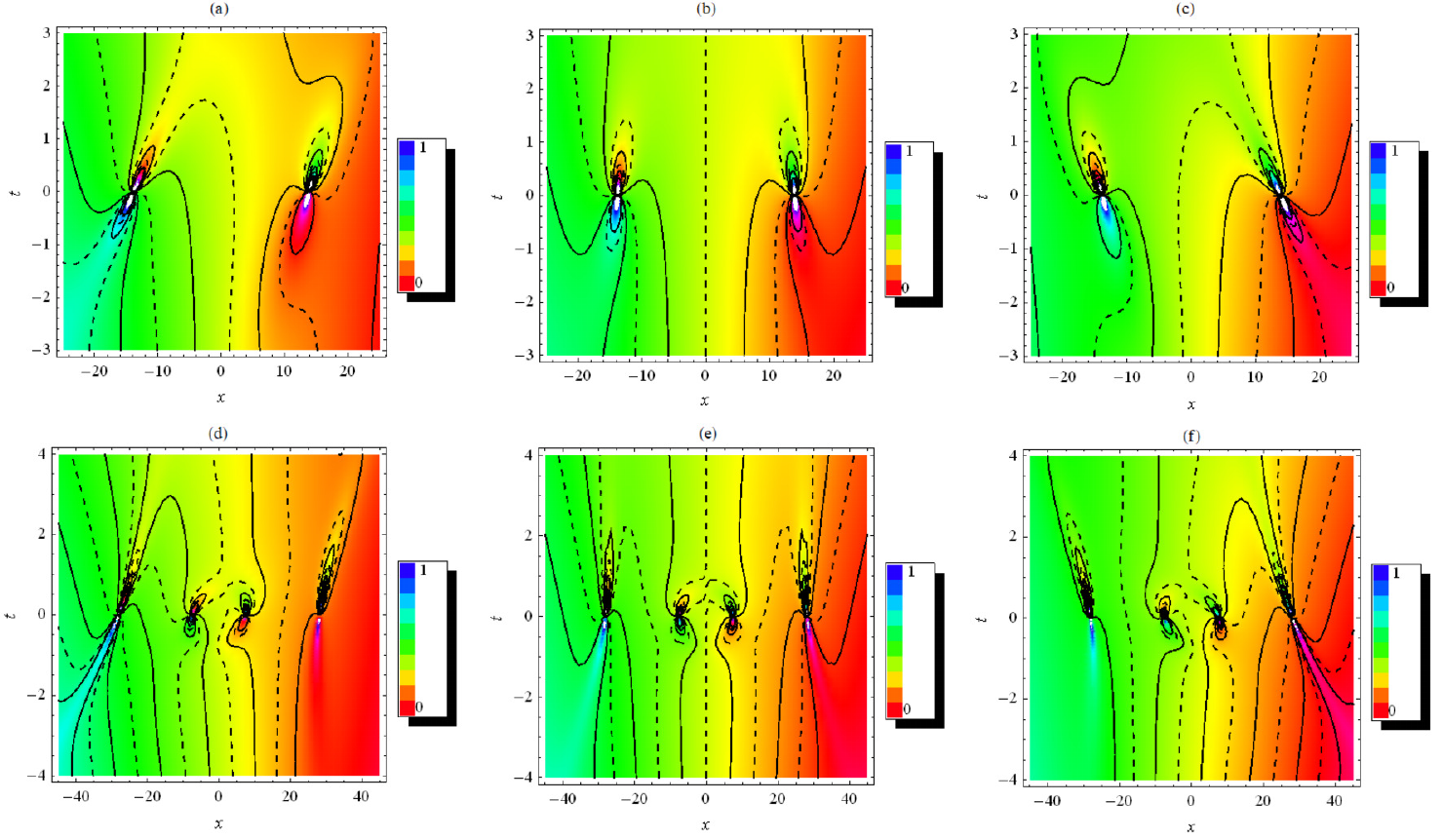}}
\caption{(Color online) The distribution of the frequency chirp (\protect\ref%
{20}) associated with the first-order RW solution (\protect\ref{18}) (top),
and frequency chirp (\protect\ref{22b}) corresponding to the second-order RW
solution (\protect\ref{22a}) (bottom). The step-like modulation format is
taken as per Eqs. (\protect\ref{step}) and (\protect\ref{astep}). The plots
are generated for $r_{0}=0.25$, $\protect\mu _{0}=0.1$, and three different
values of strength $\protect\beta $ of the delayed nonlinear response,
\textit{viz}., $\protect\beta =-0.7$ (a,d);: $\protect\beta =0$, (b,c) and $%
\protect\beta =0.7$ (e,f).}
\label{fig13}
\end{figure}

The middle and the bottom panels of Fig. 12 displays the second-order RW
solution (\ref{22a}) at different values of, respectively, $\mu _{0}$ and $%
r_{0}$. These plots show how parameters $\mu _{0}$ and $r_{0}$ from Eqs. (%
\ref{step}) and (\ref{astep}) affect the structure of the second-order RWs.
It is seen that, for small values of $\mu _{0}$ (see the middle panels) or
large values of parameter $r_{0}$ (in the bottom panels), the second-order
RW features three humps --- namely, from right to left, the main (giant)
hump, a dwarf one, and a secondary hump. The middle (bottom) panel in Fig.
12 reveals that the amplitude of the secondary hump decreases as parameter $%
\mu _{0}$ increases ($r_{0}$ decreases), and thus the hump disappears at a
critical value of $\mu _{0}$ ($r_{0}$). It is also seen from Figs. 12(d),
(e), and (f) that the amplitudes of the giant and dwarf humps increase, and
their widths decrease, with the growth of both $\mu _{0}$ and $r_{0}$. From
the middle and bottom panels of Fig. 12 we conclude that, for small values
of $\mu _{0}$, while $r_{0}$ is kept fixed, or for large values of $r_{0}$,
while $\mu _{0}$ is kept constant, the second-order RW is built of a dwarf
hump, one secondary and one giant ones, and four valleys around the center,
as seen from Figs. 12(d) and 12(i). The minimum in the four valleys is $0$,
located approximately at $\left( \pm 1.860/r_{0},0\right) $ and $\left( \pm
7.028r_{0}^{-1},0\right) $. As these points approach $\left( x,y\right)
=\left( 0,0\right) $ at $r_{0}\rightarrow +\infty $, at large values of $%
r_{0}$ the three humps fuse to form a single giant one, while the four
valleys disappear. We also note that, as the same points move to $\left( \pm
\infty ,0\right) $ at $r_{0}\rightarrow +0$, the second-order RW keeps a
single giant hump at $r_{0}$ small enough, and this RW seems like a
first-order one.

Figure 13 displays, for $r_{0}=0.25$, $\mu _{0}=0.1$, and three different
values of $\beta $, the distribution of the frequency chirp, given by Eq. (%
\ref{20}), for the first-order RW solution (\ref{18}) (top panels), and the
chirp, given by Eq. (\ref{22b}), for the second-order RW solution (\ref{22a}%
) (bottom panels). As in the two examples considered above, the chirp
associated with the first-order and the second-order RWs is localized in
time and space, and exhibits, respectively, two and four dark-bright
localized structures. Although strength $\beta $ of the delayed nonlinear
response does not strongly affect the shape of the frequency chirp, it
produces, as in the solutions considered above, an essential skew angle
relative to the ridge of the chirp in the anti-clockwise direction for $%
\beta >0$, and in the clockwise direction for $\beta <0$, as is clearly seen
in Fig. 13.

\section{Conclusion}

We have studied the generation of first- and second-order chirped RWs (rogue
waves) in the BEC\ model with the time-varying atomic scattering length in
the expulsive parabolic potential. The model is based on the cubic GP
equation (\ref{1}) for the mean-field wave function. By combining the
modified lens-type transformation with the phase engineering technique, the
cubic GP equation was transformed into the Kundu-Eckhaus equation with the
quintic nonlinearity and the term which represents the Raman effect in fiber
optics. We considered solutions based on the CW background which satisfies
condition (\ref{m0}) of the baseband modulational instability. The resulting
equation is integrable if strength $k(t)$ of the parabolic potential and
nonlinearity strength $a(t)$ satisfy integrability condition (\ref{4e}).
Using the known first- and second-order RW solutions of the KE equation, we
have presented explicit first- and second-order chirped RW solutions (\ref%
{18}) and (\ref{22a}), along with the corresponding expressions (\ref{20})
and (\ref{22b}) for the local chirp. Then, we have identified the first- and
second-order chirped RWs in the model with (i) the exponentially
time-varying atomic scattering length (Eq. (\ref{exp})), (ii) time-periodic
modulation of the nonlinearity, and (iii) the stepwise temporal modulation
based on Eqs. (\ref{step}) and (\ref{astep}). In each case, the effects
produced by strength $\beta $ of the delayed nonlinear response on the
chirped RWs are analyzed. This parameter affects the spatial location of the
humps in the first- and second-order RWs, whereas the amplitudes of the
humps and the time of their appearance remain unaltered. More interestingly,
we have found that the first-order (second-order) RWs involve a frequency
chirp that is localized in time and space. Moreover, we have found that the
chirp of the first- and second-order RW exhibits, respectively, two or four
dark-bright localized structures.

We have also studied in detail characteristics of the constructed RWs in
terms of the time-dependent parameters $a(t)$ and $k(t)$. The results
demonstrate how these parameters affect the first- and second-order RWs. It
is shown that they can be used to manage the evolution of the RWs. We have
also observed that the behavior of the RW's background changes, depending on
the temporal modulation of $a(t)$ and $k(t)$. The results of this work
suggest possibilities to manipulate RWs experimentally in BEC with the
atomic scattering length modulated in time by means of the FR (the Feshbach
resonance).\bigskip\

\textbf{Compliance with ethical standards} \bigskip

\textbf{Conflict of interest:} The authors declare that they have no
conflict of interest. \bigskip

\section*{Acknowledgment}

The work of E.K. is supported, in part, by the Initiative of the President
of the Chinese Academy of Sciences for Visiting Scientists (PIFI) under
Grants No. 2020VMA0040, the National Key R\&D Program of China under grants
No. 2016YFA0301500, NSFC under grants Nos.11434015, 61227902. W.-M.L.'s work
is supported by the National Key R\&D Program of China under grants No.
2016YFA0301500, NSFC under grants Nos.11434015, 61227902. The work of B.A.M.
is supported, in part, by Israel Science Foundation through grant No.
1286/17.

\end{document}